\documentclass[preprint,12pt]{elsarticle}

\usepackage{amssymb}
\biboptions{sort&compress}
\usepackage{amsthm}
\usepackage{graphicx}
\usepackage{multirow}
\usepackage{array}
\usepackage{booktabs}
\usepackage{threeparttable}
\usepackage{makecell}

\newcommand{\PreserveBackslash}[1]{\let\temp=\\#1\let\\=\temp}
\newcolumntype{C}[1]{>{\PreserveBackslash\centering}p{#1}}
\newcolumntype{R}[1]{>{\PreserveBackslash\raggedleft}p{#1}}
\newcolumntype{L}[1]{>{\PreserveBackslash\raggedright}p{#1}}

\journal{Nuclear Physics B}

\begin{document}

\begin{frontmatter}



\title{Validity of equivalent photon spectra and the photoproduction processes in p-p collisions}


\author[1,3]{Zhi-Lei Ma\corref{cor1}}
\ead{mzl0197@ynu.edu.cn}

\author[2]{Zhun Lu}
\ead{zhunlu@seu.edu.cn}
\author[3]{Li Zhang\corref{cor1}}
\ead{lizhang@ynu.edu.cn}

\cortext[cor1]{Corresponding author}

\affiliation[1]{organization={Department of Physics, Yunnan University},
            city={Kunming},
            postcode={650091},
            country={China}}

\affiliation[2]{organization={School of Physics, Southeast University},
            city={Nanjing},
            postcode={211189},
            country={China}}

\affiliation[3]{organization={Department of Astronomy, Key Laboratory of Astroparticle Physics of Yunnan Province, Yunnan University},
            city={Kunming},
            postcode={650091},
            country={China}}

\begin{abstract}
Through a consistent analysis of the terms neglected in going from the accurate expression to the one of Weizs\"{a}cker-Williams approximation (WWA), the validity of equivalent photon spectra is studied, and a modified photon flux of proton is also derived.
We take the photoproductions of photons and dileptons as examples, to provide the comparison between the exact results and the ones based on various photon fluxes.
We present the results for the distributions in $Q^{2}$ (virtuality of photons), $y$ and $p_{T}$, the total cross sections are also estimated.
The numerical results show that the modified equivalent photon spectrum reproduces the exact result within less than one percent. And the corrections of photoproduction processes to the dileptons and photons productions are about $20\%$.
\end{abstract}


\end{frontmatter}


\section{Introduction}
\label{Introduction}

The central idea of equivalent photon approximation was originally pointed out in 1924 by Fermi~\cite{Fermi:1924tc}.
According to the fact that a fast moving charged particle carries electric fields that point radially outward and magnetic fields circling it, the field at a point some distance away from the trajectory of the particle resembles that of a real photon, Fermi replaced the electromagnetic fields from a fast-moving charged particle with an equivalent flux of photon.
The number of photons with energy $\omega$, $n(\omega)$, is given by the Fourier transform of the time-dependent electromagnetic field \cite{Phys.Rev._51_1037, Dalitz:1957dd, Nucl.Phys._23_1295}.
Therefore, the cross section of the electromagnetic interaction is approximated by the convolution of the photon flux with the relevant real photoproduction cross section.
A decade later, in order to simplify calculations of processes involving relativistic collisions of charged particles, Weizs\"{a}cker and Williams independently
conceived the technique now known as the Weizs\"{a}cker-Williams approximation (WWA) \cite{vonWeizsacker:1934nji}.
An essential advantage of WWA consists in the fact that, when using it, it is sufficient to obtain the photo-absorption cross section on the mass shell only.
Details of its off mass-shell behavior are not essential.
Subsequently, WWA as a useful technique was substantiated and successfully applied, for instance, to two-photon processes for particle production, photoproduction mechanism, meson production in electron-nucleon collisions, the determination of the nuclear parton distributions, and small-$x$ physics~\cite{Phys. Rev._104_211, Zhu:2015via, Zhu:2015qoz, Fu:2011zzf, Fu:2011zzm, Brown:1973qj, Brodsky:1971ud,  Budnev:1974de, Chin.Phys.C_36_721, Yu:2017rfi, Yu:2015kva, Yu:2017pot, Drees:1989vq, Drees:1988pp, Frixione:1993yw, Nystrand:2004vn, Nystrand:2006gi,Ma:2021jes, Winther:1979zz}.
On the other hand, WWA has found application beyond the realms of QED, such as the equivalent pion method which describes the subthreshold pion production in nucleus-nucleus collision~\cite{Pirner:1980rn};
the nuclear Weizs\"{a}cker-Williams method which describes excitation processes induced by the nuclear interaction in peripheral collisions of heavy ions \cite{Feshbach:1976uu};
a non-Abelian Weizs\"{a}cker-Williams method describing the boosted gluon distribution functions in nucleus-nucleus collision~\cite{McLerran:1994vd}.

Although the tremendous successes have been achieved, the discussion about the accuracy of WWA and its applicability range is still insufficient.
A number of widely used equivalent photon fluxes are proposed beyond the WWA validity range, and some imprecise statements are given~\cite{Drees:1989vq, Drees:1988pp, Frixione:1993yw, Yu:2015kva, Yu:2017pot, Yu:2017rfi, Zhu:2015via, Zhu:2015qoz, Fu:2011zzf, Chin.Phys.C_36_721, Winther:1979zz}.
In Ref.\cite{Drees:1988pp}, Drees and Zeppenfeld studied the production of supersymmetric particles in elastic $ep$ collisions, in order to simplify the calculations from the exact computation, they derived a equivalent photon spectrum of high energy protons by using the electric dipole form factor.
They also used a exponential form factor of heavy nuclei to derive a photon spectrum for lead \cite{Drees:1989vq}.
In Ref.\cite{Nystrand:2004vn}, Nystrand derived a modified photon spectrum by considering the $m_{p}^{2}$ term which was neglected by Drees and Zeppenfeld and employed it to study the electromagnetic interactions in nucleus-nucleus collisions, as well as the ultra-peripheral collisions of heavy ions at RHIC and LHC in Ref.\cite{Nystrand:2006gi}.
The effect of including the magnetic dipole moment and the corresponding magnetic form factor of the proton has been investigated by Kniehl in Ref.\cite{Kniehl:1990iv}, where a modified WWA is formulated for the elastic $ep$ scattering.
The authors in Refs.\cite{Cahn:1990jk, Baur:1990fx} defined a semiclassical impact parameter dependent equivalent photon spectrum, which excludes the hadronic interaction easily.
Brodsky and Drees achieved two different forms of photon spectra inside a quark in Refs.\cite{Brodsky:1971ud, Terazawa:1973tb}, which are employed to study the two-photon mechanism of particle production at high energy colliding beams.
The integration of the equivalent photon spectra in all of theses above works are performed over the entire kinematically allowed region, where the WWA errors are included.

Since the equivalent photon spectrum plays the fundamental role in the calculations of photoproduction processes, in present work, we take photoproductions of photons and dileptons  as examples to discuss the validity of several widely used equivalent photon spectra mentioned above.
We present the comparison between the WWA results and the exact ones to analyse the source of WWA errors.
As a result, we derive a modified photon flux of proton and apply it to the calculations of $p_{T}$ dependent cross sections.


\section{General formalism}
\label{General formalism}

A consistent analysis of the terms neglected in going from the accurate expression of diagram Fig.~\ref{fig:feyn}(a) to the WWA one permits in a natural manner to estimate the errors of equivalent photon spectra. This can be performed in a general form for every reaction.
In the process described in Fig.~\ref{fig:feyn}, the virtual photons radiated from the projectile $\alpha$ are off mass shell and no longer transversely polarized.
Therefore, the accurate form of the cross section can be derived based on the expansion of the proton or quark tensor (multiplied by $Q^{-2}$) by using the transverse and longitudinal polarization operators.

\subsection{The accurate expression of cross section}
\label{The accurate expression}
~~~~~~~~~~

\begin{figure*}
\centering
\includegraphics[width=0.42\columnwidth]{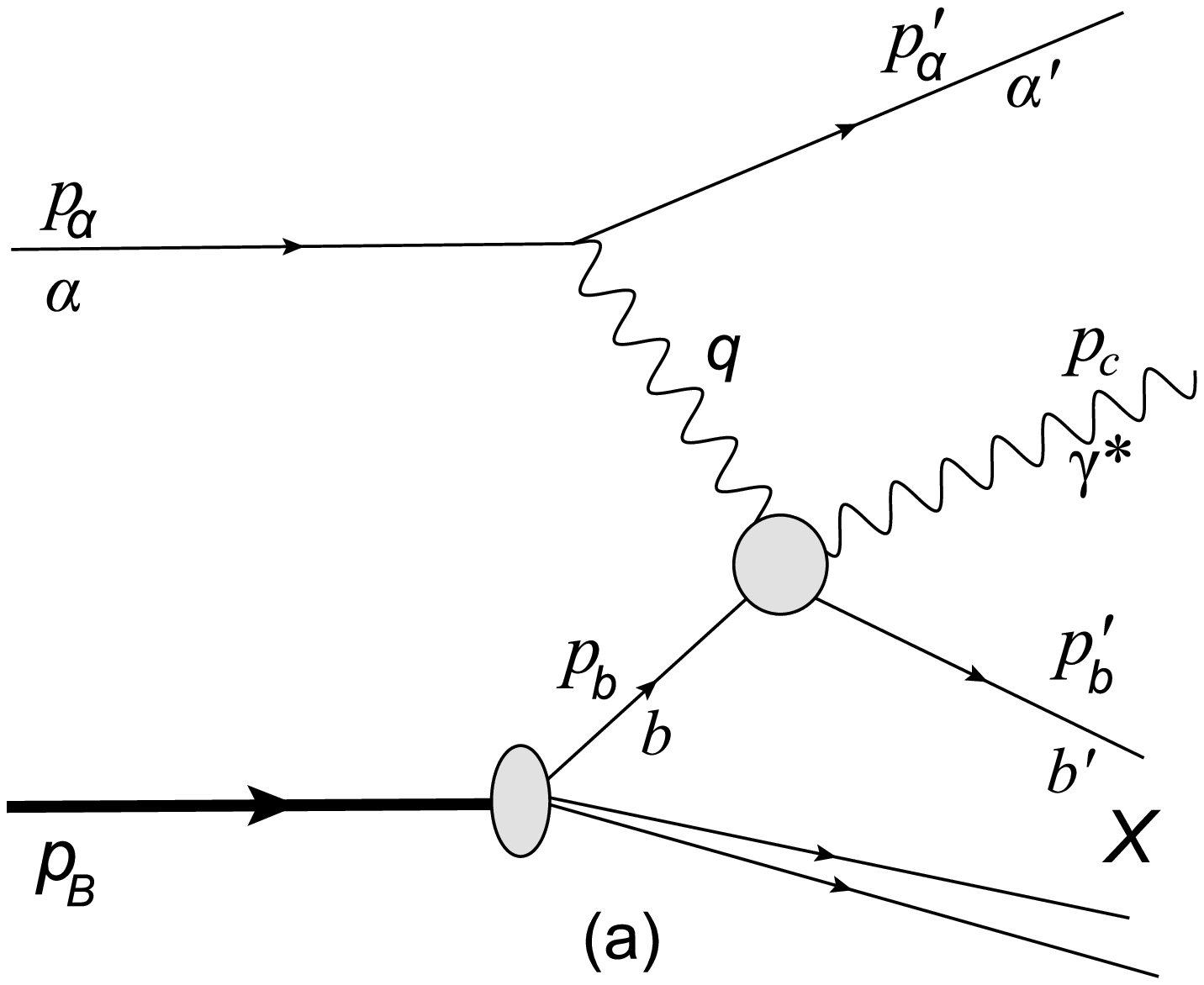}\hspace{25mm}
\includegraphics[width=0.30\columnwidth]{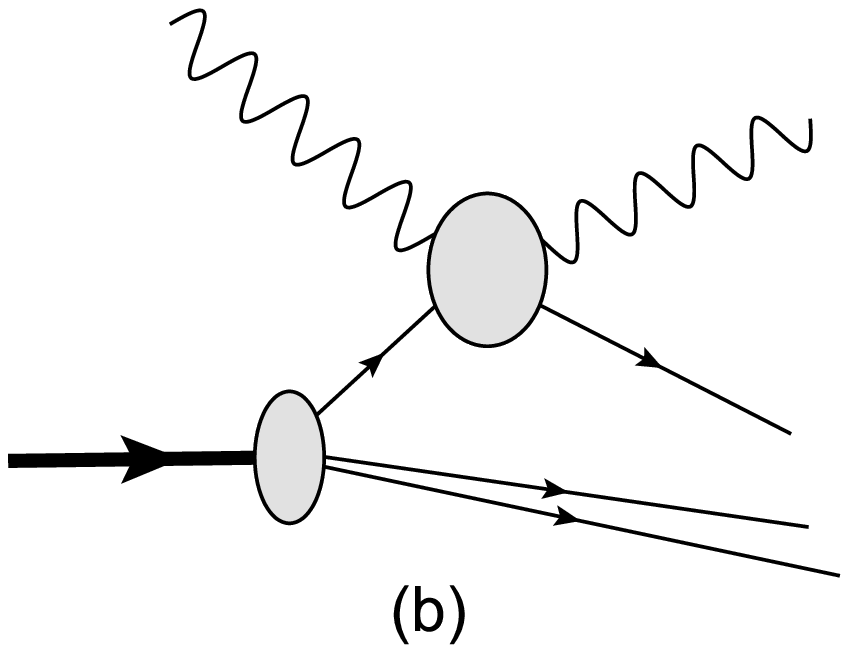}
\caption{(a): The general photoproduction processes, the virtual photon emitted from the projectile $\alpha$ interacts with parton $b$ in proton $B$. $X$ is the sum of residue of $B$ after scattering with photon. (b): photo-absorption.}
\label{fig:feyn}
\end{figure*}

The general form of the cross section for $\alpha p\rightarrow\alpha\gamma^{*}X$ described in Fig.~\ref{fig:feyn} can be written as
\begin{eqnarray}
d\sigma(\alpha+p\rightarrow \alpha+\gamma^{*}+X)
=\sum_{b}\int dx_{b}f_{b/p}(x_{b},\mu_{b}^{2})d\sigma(\alpha+b\rightarrow \alpha+\gamma^{*}+b),
\end{eqnarray}
where $x_{b}=p_{b}/p_{B}$ is the momentum fraction of parton $b$ struck by the virtual photon, $f_{b/p}(x_{b},\mu_{b}^{2})$ is the parton distribution function of massless parton $b$ in proton $B$, and $\mu_{b}$ is the factorized scale.
$d\sigma(\alpha+b\rightarrow \alpha+\gamma^{*}+b)$ is the differential cross section in the parton level and can be presented as
\begin{eqnarray}\label{dab.Gen1}
d\sigma(\alpha+b\rightarrow \alpha+\gamma^{*}+b)
=\frac{|\mathcal{M}_{\alpha b}|^{2}}{4\left[(p_{\alpha}\cdot p_{b})^{2}-p_{\alpha}^{2}p_{b}^{2}\right]^{1/2}}d\textrm{PS}_{3}(p_{\alpha}+p_{b};p_{\alpha}',p_{c},p_{b}'),
\end{eqnarray}
where we employ the short-hand notation
\begin{eqnarray}\label{dab.Gen1}
d\textrm{PS}_{n}(P;p_{1},...,p_{n})=(2\pi)^{4}\delta^{4}\left(P-\sum_{i=1}^{n}p_{i}\right)\prod_{i=1}^{n}\frac{d^{3}p_{i}}{(2\pi)^{3}2E_{i}},
\end{eqnarray}
for the Lorentz invariant N-particle phase-space element.
By decomposing the squared scattering amplitude as $|\mathcal{M}_{\alpha b}|^{2}=4\pi\alpha_{\mathrm{em}} e_{\alpha}^{2}\rho^{\mu\nu}T_{\mu\nu}/Q^{2}$ and rearranging $d\textrm{PS}_{n}$, the cross section of subprocess $\alpha+b\rightarrow \alpha+\gamma^{*}+b$ can be further expressed as follows
\begin{eqnarray}\label{dab.Gen2}
&&\!\!\!\!d\sigma(\alpha+b\rightarrow \alpha+\gamma^{*}+b)\nonumber\\
=\!\!\!\!&&\!\!\!\!4\pi e_{\alpha}^{2}\alpha_{\mathrm{em}}\frac{\rho^{\mu\nu}}{Q^2}\frac{d^{3}p'_{\alpha}}{(2\pi)^{3}2E'_{\alpha}}\left[\frac{(q\cdot p_{b})^{2}-q^{2}p_{b}^{2}}{(p_{\alpha}\cdot p_{b})^{2}-p_{\alpha}^{2}p_{b}^{2}}\right]^{1/2}T_{\mu\nu}\frac{d\mathrm{PS}_{2}(q+p_{b};p_{c},p_{b}')}{4\hat{p}_{\mathrm{CM}}\sqrt{\hat{s}}},
\end{eqnarray}
where $e_{\alpha}$ is the charge of the projectile $\alpha$, $\alpha_{\mathrm{em}}=1/137$ is the fine structure constant, and
\begin{eqnarray}\label{lambda}
\left[\frac{(q\cdot p_{b})^{2}-q^{2}p_{b}^{2}}{(p_{\alpha}\cdot p_{b})^{2}-p_{\alpha}^{2}p_{b}^{2}}\right]^{1/2}
=\frac{\hat{p}_{\mathrm{CM}}\sqrt{\hat{s}}}{p_{\mathrm{CM}}\sqrt{s_{\alpha b}}},
\end{eqnarray}
$s_{\alpha b}=(p_{\alpha}+p_{b})^{2}$ and $\hat{s}=(q+p_{b})^{2}$ are the energy square of $\alpha b$ and $\gamma^{*}b$ CM frames, respectively.
$p_{\mathrm{CM}}$ and $\hat{p}_{\mathrm{CM}}$ are the momenta of corresponding CM frames.
$T_{\mu\nu}$ is the amplitude of reaction $\gamma^{*}+b\rightarrow \gamma^{*}+b$, and the $\rho^{\mu\nu}$ quantity is the density matrix of the virtual photon produced by the projectile $\alpha$,
\begin{eqnarray}\label{Rou.Gen}
\rho^{\mu\nu}=(-g^{\mu\nu}+\frac{q^{\mu}q^{\nu}}{q^{2}})F_{2}(Q^{2})-\frac{(2p_{\alpha}-q)^{\mu}(2p_{\alpha}-q)^{\nu}}{q^{2}}F_{1}(Q^{2}),
\end{eqnarray}
$F_{1}(Q^{2})$ and $F_{2}(Q^{2})$ are the general expressions for the form factors of projectile.

It is convenient to use the following linear combinations~\cite{Budnev:1974de}
\begin{eqnarray}\label{Line.QR}
Q^{\mu}\!\!&=&\!\!\sqrt{\frac{-q^{2}}{(q\cdot p_{b})^{2}-q^{2}p_{b}^{2}}}(p_{b}-q\frac{q\cdot p_{b}}{q^{2}})^{\mu},\nonumber\\
R^{\mu\nu}\!\!&=&\!\!-g^{\mu\nu}+\frac{(q\cdot p_{b})(q^{\mu}p_{b}^{\nu}+q^{\nu}p_{b}^{\mu})-q^{2}p_{b}^{\mu}p_{b}^{\nu}
-p_{b}^{2}q^{\mu}q^{\nu}}{(q\cdot p_{b})^{2}-q^{2}p_{b}^{2}},
\end{eqnarray}
they satisfy the relations: $q_{\mu}Q^{\mu}=q_{\mu}R^{\mu\nu}=0$, $Q^{\mu}Q_{\mu}=1$, thus $\rho^{\mu\nu}$ can be expanded to
\begin{eqnarray}\label{Rou.Exp}
\rho^{\mu\nu}=\rho^{00}Q^{\mu}Q^{\nu}+\rho^{++}R^{\mu\nu},
\end{eqnarray}
where
$\rho^{++}=R^{\mu\nu}\rho_{\mu\nu}/2$, $\rho^{00}=Q^{\mu}Q^{\nu}\rho_{\mu\nu}$.
With the relations
\begin{eqnarray}\label{subab.TL}
2d\sigma_{T}(\gamma^{*}+b\rightarrow \gamma^{*}+b)
\!\!&=&\!\!R^{\mu\nu}T_{\mu\nu}\frac{d\mathrm{PS}_{2}(q+p_{b};p_{c},p_{b}')}{4\hat{p}_{\mathrm{CM}}\sqrt{\hat{s}}},\nonumber\\
d\sigma_{L}(\gamma^{*}+b\rightarrow \gamma^{*}+b)
\!\!&=&\!\!Q^{\mu}Q^{\nu}T_{\mu\nu}\frac{d\mathrm{PS}_{2}(q+p_{b};p_{c},p_{b}')}{4\hat{p}_{\mathrm{CM}}\sqrt{\hat{s}}},
\end{eqnarray}
the differential cross section of subprocess $\alpha+b\rightarrow \alpha+\gamma^{*}+b$ can finally be expressed as
\begin{eqnarray}\label{dabTL}
&&\!\!\!\!d\sigma(\alpha+b\rightarrow \alpha+\gamma^{*}+b)\nonumber\\
=\!\!\!\!&&\!\!\!\!\frac{e_{\alpha}^{2}\alpha_{\mathrm{em}}}{2\pi^{2}Q^{2}}\left[\rho^{++}d\sigma_{T}(\gamma^{*}+b\rightarrow \gamma^{*}+b)+\frac{\rho^{00}}{2}d\sigma_{L}(\gamma^{*}+b\rightarrow \gamma^{*}+b)\right]\nonumber\\
\!\!\!\!&&\!\!\!\!\times\frac{\hat{p}_{\mathrm{CM}}\sqrt{\hat{s}}}{p_{\mathrm{CM}}\sqrt{s_{\alpha b}}}\frac{d^{3}p'_{\alpha}}{E'_{\alpha}},
\end{eqnarray}
and
\begin{eqnarray}\label{Rouzz00}
\rho^{++}\!\!&=&\!\!F_{2}(Q^{2})+\frac{1}{2}\left[\frac{(2-y)^{2}}{y^{2}+Q^{2}m_{b}^{2}/(p_{\alpha}\cdot p_{b})^{2}}-\frac{4m_{\alpha}^{2}}{Q^{2}}-1\right]F_{1}(Q^{2}),\nonumber\\
\rho^{00}\!\!&=&\!\!-F_{2}(Q^{2})+\frac{(2-y)^{2}}{y^{2}+Q^{2}m_{b}^{2}/(p_{\alpha}\cdot p_{b})^{2}}F_{1}(Q^{2}).
\end{eqnarray}
Here $Q^{2}=-q^{2}$ and $d\sigma_{T (L)}/d\hat{t}$ represents the transverse (longitudinal) cross section of subprocess $\gamma^{*}+b\rightarrow\gamma^{*}+b$, its analytical expression can be found in Ref.\cite{Ma:2019mwr}.

\begin{figure*}
\centering
\includegraphics[width=0.35\columnwidth]{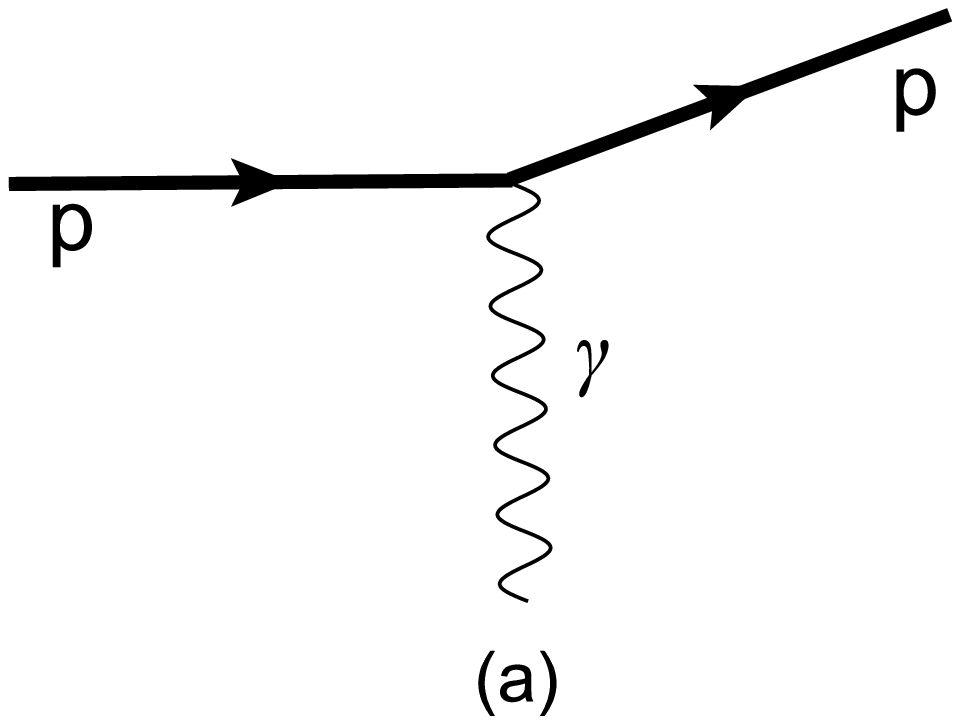}\hspace{20mm}
\includegraphics[width=0.43\columnwidth]{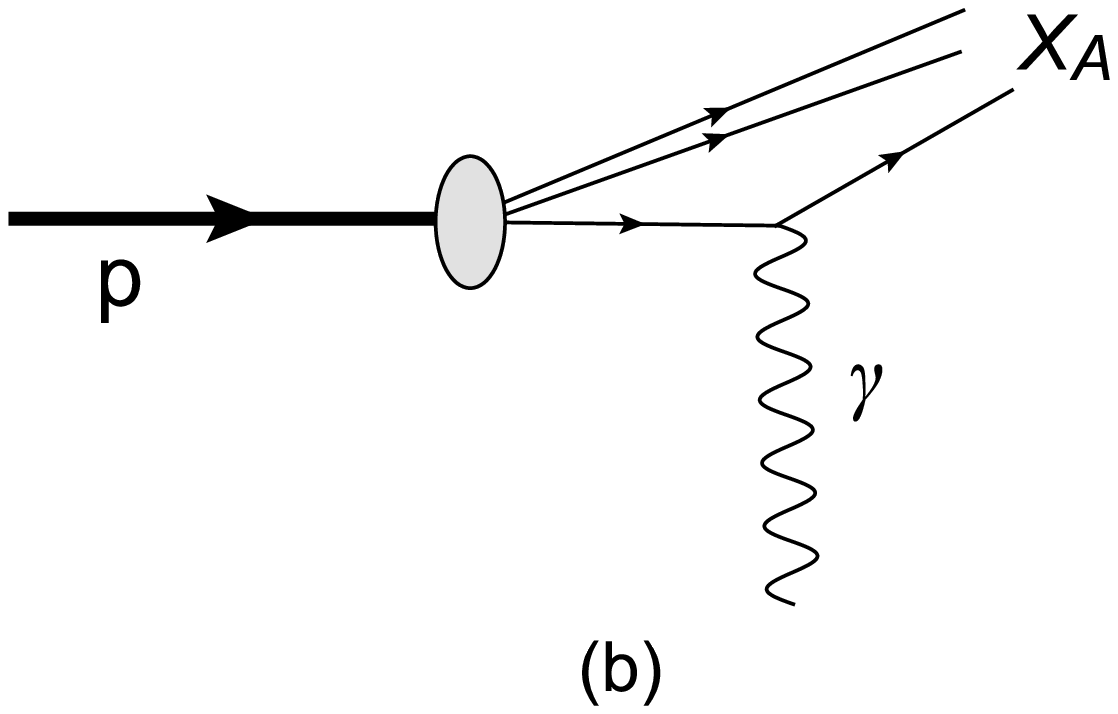}
\caption{(a): Coherent emission, virtual photon is radiated coherently by the whole proton which remains intact after scattering.
(b): Incoherent emission, virtual photon is radiated incoherently by the quarks inside proton which is allowed break up after scattering.}
\label{fig:photoemi.}
\end{figure*}

We have to deal with two types of photon emission mechanisms which are described in Fig.~\ref{fig:photoemi.}: coherent emission (coh.) and incoherent emission (incoh.).
In the first type, the virtual photon is radiated coherently by the whole proton which remains intact after photon emission.
In the second type, the virtual photon is radiated incoherently by the quarks inside the proton, and the proton will dissociate or excite after photon emission.
In the case of coherent reactions in Fig.~\ref{fig:photoemi.}(a), the projectile $\alpha$ is proton: $m_{\alpha}=m_{p}$, and thus the general expressions of form factor $F_{1}(Q^{2})$ and $F_{2}(Q^{2})$ turn into the elastic proton form factors accordingly. Then the density matrix of the virtual photon in Eq.~(\ref{Rouzz00}) reads
\begin{eqnarray}\label{Rouzz00.coh.}
\rho^{++}_{\mathrm{coh}.}\!\!&=&\!\!G_{\mathrm{E}}^{2}(Q^{2})\left[7.78-\frac{1}{2}\left(\frac{4m_{\alpha}^{2}+Q^{2}}{Q^{2}}
-\frac{(2-y)^{2}}{y^{2}}\right)\frac{4m_{p}^{2}+7.78Q^{2}}{4m_{p}^{2}+Q^{2}}\right],\nonumber\\
\rho^{00}_{\mathrm{coh}.}\!\!&=&\!\!G_{\mathrm{E}}^{2}(Q^{2})\left[-7.78+\frac{(2-y)^{2}}{y^{2}}\frac{4m_{p}^{2}+7.78Q^{2}}{4m_{p}^{2}+Q^{2}}\right],
\end{eqnarray}
where the electric form factor of proton can be parameterized by a dipole form
\begin{eqnarray}\label{GE}
G_{\mathrm{E}}(Q^{2})=\frac{1}{(1+Q^{2}/0.71~\mathrm{GeV}^{2})^{2}}.
\end{eqnarray}

In the case of incoherent reactions in Fig.~\ref{fig:photoemi.}(b), the projectile $\alpha$ is the quark inside the proton: $m_{\alpha}=m_{q}=0$.
Besides, the Martin-Ryskin method is adopted for the calculations \cite{Martin:2014nqa}, where the square of the form factor is used as the coherent probability or weighting factor (WF): $\omega_{c}=G_{\mathrm{E}}^{2}(Q^2)$, and in order to avoid double counting, the "remaining" probability has to be considered in the incoherent case: $1-\omega_{c}=1-G_{\mathrm{E}}^{2}(Q^2)$.
Thus, we have $F_{1}(Q^{2})=F_{2}(Q^{2})=1-G_{\mathrm{E}}^{2}(Q^{2})$, and the density matrix of the virtual photon in Eq.~(\ref{Rouzz00}) becomes
\begin{eqnarray}\label{Rouzz00.incoh.}
&&\rho^{++}_{\mathrm{incoh.}}=\left[1-G_{\mathrm{E}}^{2}(Q^{2})\right]\left[\frac{1}{2}+\frac{(y-2)^{2}}{2y^{2}}\right],\nonumber\\
&&\rho^{00}_{\mathrm{incoh.}}=\left[1-G_{\mathrm{E}}^{2}(Q^{2})\right]\left[\frac{(y-2)^{2}}{y^{2}}-1\right].
\end{eqnarray}
~~~~~~~~~~

\subsection{The equivalent photon spectrum}
\label{The eps}
~~~~~~~~~~

The connection between the process in Fig.~\ref{fig:feyn}(a) and the one in (b) is evident.
The WWA consists in ignoring the fact that the photon in this photo-absorption amplitude [Fig.~\ref{fig:feyn}(b)] is off mass shell and no longer transversely polarized from real photo-absorption.
As a result, the photoprodution process described in Fig.~\ref{fig:feyn}(a) can be factorized in terms of the real photo-absorption cross section and the equivalent photon spectrum.
When switching to the approximate formulae of WWA, two simplifications should be performed in accurate expression [Eq.~(\ref{dabTL})].
First, the scalar photon contribution $\sigma_{L}$ is neglected; secondly, the term of $\sigma_{T}$ is substituted by its on-shell value.
This provides us a powerful approach for comparing our results with the WWA ones to study the features of the equivalent photon spectra which are widely applied in the literatures.

Taking $Q^{2}\rightarrow0$, the linear combinations in Eq.~(\ref{Line.QR}) can reduce to
\begin{eqnarray}\label{epsi.TL}
&&\lim_{Q^{2}\rightarrow0}Q^{\mu}Q^{\nu}=\frac{q^{\mu}q^{\nu}}{q^{2}},\nonumber\\
&&\lim_{Q^{2}\rightarrow0}R^{\mu\nu}=-g^{\mu\nu}+\frac{(q^{\mu}p_{b}^{\nu}+q^{\nu}p_{b}^{\mu})}{q\cdot p_{b}}.
\end{eqnarray}
According to gauge invariance $q^{\mu}T_{\mu\nu}=0$, Eq. (\ref{dabTL}) is simplified to:
\begin{eqnarray}\label{dWWA.Gen.}
&&\!\!\!\!\lim_{Q^{2}\rightarrow0}d\sigma(\alpha+b\rightarrow\alpha+\gamma^{*}+b)\nonumber\\
=\!\!\!\!&&\!\!\!\!\left(e_{\alpha}^{2}\frac{\alpha_{\mathrm{em}}}{2\pi^{2}}\frac{y\rho^{++}}{Q^{2}}\frac{d^{3}p'_{\alpha}}{E'_{\alpha}}\right)
\frac{\hat{p}_{\mathrm{CM}}\sqrt{\hat{s}}}{yp_{\mathrm{CM}}\sqrt{s_{\alpha b}}}d\sigma_{T}(\gamma^{*}+b\rightarrow \gamma^{*}+b)\bigg|_{Q^{2}=0}\nonumber\\
=\!\!\!\!&&\!\!\!\!\left[e_{\alpha}^{2}\frac{\alpha_{\mathrm{em}}}{2\pi}(y\rho^{++})dy\frac{dQ^{2}}{Q^{2}}\right]d\sigma_{T}(\gamma^{*}+b\rightarrow \gamma^{*}+b)\bigg|_{Q^{2}=0}\nonumber\\
=\!\!\!\!&&\!\!\!\!dn_{\gamma}d\sigma_{T}(\gamma^{*}+b\rightarrow \gamma^{*}+b)\bigg|_{Q^{2}=0},
\end{eqnarray}
where $y=(q\cdot p_{b})/(p_{\alpha}\cdot p_{b})$ is the relative energy loss of the projectile $\alpha$, and the general form of the equivalent photon flux $f(y)$ reads
\begin{eqnarray}\label{fgamma.Gen.}
f(y)\!\!\!\!&=&\!\!\!\!\frac{dn_{\gamma}}{dy}=\int e_{\alpha}^{2}\frac{\alpha_{\mathrm{em}}}{2\pi}y\rho^{++}\frac{dQ^{2}}{Q^{2}}\nonumber\\
\!\!\!\!&=&\!\!\!\!e_{\alpha}^{2}\frac{\alpha_{\mathrm{em}}}{2\pi}y\int\frac{dQ^{2}}{Q^{2}}\left[F_{2}(Q^{2})
+\left(\frac{2(1-y)}{y^{2}}-\frac{2m_{\alpha}^{2}}{Q^{2}}\right)F_{1}(Q^{2})\right].
\end{eqnarray}

In the case of coherent photon emission of proton, we derived a photon flux function from Eq.~(\ref{fgamma.Gen.}).
By neglecting the contributions from the magnetic form factor and adopting the dipole form of electric form factor of proton: $F_{1}(Q^{2})=F_{2}(Q^{2})=G_{\mathrm{E}}^{2}(Q^{2})$, and employing the coherent condition $Q^{2}\leq 1/R^{2}_{A}$ ($R_{A}=A^{1/3}1.2\ \textrm{fm}$ is the size of the nucleus), one obtains with $a=2m_{p}^{2}/Q^{2}_{\mathrm{max}}$ and $b=2m_{p}^{2}/0.71=2.48$,
\begin{eqnarray}\label{fgamma.MD.}
f_{\mathrm{MD}}(y)\!\!\!\!&=&\!\!\!\!\frac{\alpha_{\mathrm{em}}}{2\pi}y\left[a-2x+(2x+c_{1})d_{1}+(2x+c_{2})d_{2}\right.\nonumber\\
&&\!\!\!\!+\left.(3x+c_{3})d_{3}+(2x+c_{4})d_{4}\right],
\end{eqnarray}
where $x$ depends on $y$,
\begin{eqnarray}\label{x}
x=-\frac{1}{y}+\frac{1}{y^{2}}.
\end{eqnarray}
The factors $c_{i}$ and $d_{i}\ (i=1,...,4)$ in Eq.~(\ref{fgamma.MD.}) have the forms
\begin{eqnarray}\label{x}
&&c_{1}=1+4b\approx 10.92,\ \ \  d_{1}=\ln\frac{A}{A'},\nonumber\\
&&c_{2}=1+2b\approx 5.96,\ \ \ \ \  d_{2}=3(\frac{1}{A}-\frac{1}{A'}),\nonumber\\
&&c_{3}=\frac{3}{2}+2b\approx 6.46,\ \ \ \ d_{3}=-(\frac{1}{A^{2}}-\frac{1}{A'^{2}}),\nonumber\\
&&c_{4}=1+b\approx 3.48,\ \ \ \ \ \ d_{4}=\frac{1}{3}(\frac{1}{A^{3}}-\frac{1}{A'^{3}}),
\end{eqnarray}
where $A=(1+0.71~\mathrm{GeV}^{2}/Q^{2}_{\mathrm{min}})$, $A'=(1+0.71~\mathrm{GeV}^{2}/Q^{2}_{\mathrm{max}})$ and
\begin{eqnarray}\label{Q2lim.}
Q^{2}_{\mathrm{min}}\!\!\!\!&=&\!\!\!\!-2m_{\alpha}^{2}+\frac{1}{2s_{\alpha b}}\left[(s_{\alpha b}+m_{\alpha}^{2})(s_{\alpha b}-\hat{s}+m_{p}^{2})\right.\nonumber\\
\!\!\!\!&&\!\!\!\!\left.-(s_{\alpha b}-m_{\alpha}^{2})\sqrt{(s_{\alpha b}-\hat{s}+m_{\alpha}^{2})^{2}-4s_{\alpha b}m_{\alpha}^{2}}\right],\nonumber\\
Q^{2}_{\mathrm{max}}\!\!\!\!&=&\!\!\!\!1/R^{2}_{A}=0.027,
\end{eqnarray}
with $m_{\alpha}=m_{p}$.
Since the coherent condition is employed, $Q^{2}_{\mathrm{max}}$ is limited to very low value.
One should note that, since $y$ depends on $Q^{2}$, we can not set $y_{\mathrm{max}}=1$ directly in the calculation, instead $y_{\mathrm{max}}=0.16$ in this case.

Actually, the origin of various practically employed photon spectra is another plane wave form, which is given in Ref.\cite{Budnev:1974de} and can be presented as follows
\begin{eqnarray}\label{fgamma.Gen.V}
dn_{\gamma}=e_{\alpha}^{2}\frac{\alpha_{\mathrm{em}}}{\pi}\frac{dy}{y}\frac{dQ^{2}}{Q^{2}}\left[(1-y)\frac{Q^{2}-Q^{2}_{\mathrm{min}}}{Q^{2}}F_{1}(Q^{2})
+\frac{y^{2}}{2}F_{2}(Q^{2})\right],
\end{eqnarray}
this form can be also derived from Eq.~(\ref{fgamma.Gen.}) by assuming that $Q^{2}_{\mathrm{min}}=y^{2}m_{\alpha}^{2}/(1-y)$, which is the leading order (LO) term of complete expression Eq.~(\ref{Q2lim.}) in the expansion of $\mathcal{O}(m_{\alpha}^{2})$, and is only valuable when $m_{\alpha}^{2}\ll1\ \mathrm{GeV}^{2}$.
However, $m_{p}^{2}\approx 0.88\ \mathrm{GeV}^{2}$ does not satisfies this condition, this leads to about $10\%$ errors in various spectra.

In Ref.\cite{Drees:1988pp}, Drees and Zeppenfeld provided a approximate analytic form of Eq.~(\ref{fgamma.Gen.V}) which is widely used in the literatures \cite{Zhu:2015qoz,Yu:2015kva,Yu:2017rfi,Yu:2017pot,Fu:2011zzf,Chin.Phys.C_36_721}.
By taking $Q^{2}_{\mathrm{max}}\rightarrow\infty$ and setting $F_{1}(Q^{2})=F_{2}(Q^{2})=G_{\mathrm{E}}^{2}(Q^{2})$ and $Q^{2}-Q^{2}_{\mathrm{min}}\approx Q^{2}$, they obtained
\begin{eqnarray}\label{fgamma.DZ.}
f_{\mathrm{DZ}}(y)=\frac{\alpha_{\mathrm{em}}}{2\pi}\frac{1+(1-y)^{2}}{y}\left[\ln A-\frac{11}{6}+\frac{3}{A}-\frac{3}{2A^{2}}+\frac{1}{3A^{2}}\right].
\end{eqnarray}
Based on Eq.~(\ref{fgamma.DZ.}), Nystrand derived a modified photon spectrum which include the $Q^{2}_{\mathrm{min}}$ term in Eq.~(\ref{fgamma.Gen.V}) and can be presented as \cite{Nystrand:2004vn}
\begin{eqnarray}\label{fgamma.Ny.}
f_{\mathrm{Ny}}(y)=\frac{\alpha_{\mathrm{em}}}{2\pi}\frac{1+(1-y)^{2}}{y}\left[\frac{A+3}{A-1}\ln A-\frac{17}{6}-\frac{4}{3A}+\frac{1}{6A^{2}}\right].
\end{eqnarray}
In addition, the effect from including the magnetic dipole moment and the corresponding magnetic form factor of the proton has been estimated by Kniehl \cite{Kniehl:1990iv}.
The final expression $f_{\mathrm{Kn}}(y)$ (Eq.~(3.11) of \cite{Kniehl:1990iv}) is too long to include here, but will be discussed further below.

Another most important approach for photon spectrum is the semiclassical impact parameter description, which excludes the hadronic interaction easily.
The calculation of the semiclassical photon spectrum is explained in Ref.~\cite{CED}, and the result can be written as
\begin{eqnarray}\label{fgamma.SC}
f_{\mathrm{SC}}(y)=\frac{2Z^{2}\alpha_{\mathrm{em}}}{\pi}\left(\frac{c}{\upsilon}\right)^{2}\frac{1}{y}\left[\xi K_{0}K_{1}+\frac{\xi^{2}}{2}\left(\frac{\upsilon}{c}\right)^{2}\left(K^{2}_{0}-K^{2}_{1}\right)\right],
\end{eqnarray}
where $\upsilon$ is the velocity of the point charge $Ze$, $K_{0}(x)$ and $K_{1}(x)$ are the modified Bessel functions, and $\xi=b_{\mathrm{min}}m_{A}y/\upsilon$.

In the case of incoherent photon emission, the complete form of photon spectrum can be derived from Eq.~(\ref{fgamma.Gen.}) by setting $F_{1}(Q^{2})=F_{2}(Q^{2})=1-G_{\mathrm{E}}^{2}(Q^{2})$ and $m_{\alpha}=0$,
\begin{eqnarray}\label{fgamma.incoh.}
dn_{\gamma}(y)=e_{\alpha}^{2}\frac{\alpha_{\mathrm{em}}}{2\pi}dy\frac{dQ^{2}}{Q^{2}}\frac{1+(1-y)^{2}}{y}\left[1-G_{\mathrm{E}}^{2}(Q^{2})\right].
\end{eqnarray}
Actually, another approximate form of above equation is often used in practical calculations \cite{Zhu:2015qoz,Yu:2015kva,Yu:2017rfi,Yu:2017pot,Fu:2011zzf,Chin.Phys.C_36_721,Fu:2011zzm}, which neglect $G_{\mathrm{E}}^{2}(Q^{2})$ term of Eq.~(\ref{fgamma.incoh.}) and take $Q_{\mathrm{min}}^{2}=1~\mathrm{GeV}^{2}$ and $Q^{2}_{\mathrm{max}}=\hat{s}/4$,
\begin{eqnarray}\label{fgamma.incohI.}
f_{\mathrm{incoh}}(y)=e_{\alpha}^{2}\frac{\alpha_{\mathrm{em}}}{2\pi}\frac{1+(1-y)^{2}}{y}\ln\frac{Q_{\mathrm{max}}^{2}}{Q_{\mathrm{min}}^{2}}.
\end{eqnarray}

Finally, another important form of equivalent photon spectrum of parton is given by Brodsky, Kinoshita and Terazawa in Ref.~\cite{Brodsky:1971ud}, which can be expressed as
\begin{eqnarray}\label{fgamma.incohBKT}
&&\!\!\!\!\!\!f_{\mathrm{BKT}}(y)\nonumber\\
=&&\!\!\!\!\!\!e_{\alpha}^{2}\frac{\alpha_{\mathrm{em}}}{\pi}\Bigg\{\frac{1+(1-y)^{2}}{y}\left(\ln\frac{E}{m}-\frac{1}{2}\right)\nonumber\\
&&\!\!\!\!\!\!+\frac{y}{2}\left[\ln(\frac{2}{y}-2)+1\right]+\frac{(2-y)^{2}}{2y}\ln(\frac{2-2y}{2-y})\Bigg\}.
\end{eqnarray}
~~~~~~~~~~

\section{The $Q^{2}$, $y$ and $p_{T}$ distributions for the photoproductions of photons and dileptons}
\label{Distributions}
~~~~~~~~~~

The equivalent photon spectrum is the key function in the calculations of photoproduction processes.
Since photons and dileptons are the ideal probes of strong interaction matter (quark-gluon plasma, QGP), its photoproduction processes have received many studies within WWA.
We choose the photoproductions of photons and dileptons as examples, to discuss the features of the spectra mentioned above and illustrate the virtue of the photon flux Eq.~(\ref{fgamma.MD.}) by comparing their results with the accurate one in Eq.~(\ref{dabTL}).
In present section, we give the corresponding cross sections.
There are two types of photon contributions that should be considered: direct-photon and resolved-photon contributions~\cite{Ma:2018zzq}.
For the direct-photon process, the high-energy photon, emitted from the projectile $\alpha$, interacts with the partons $b$ of target proton $B$ by the interactions of quark-photon Compton scattering.
For the resolved-photon process, the high-energy photon can fluctuate into a color singlet state with multiple $q\bar{q}$ pairs and gluons.
Due to this fluctuation, the photon interacts with the partons in $B$ like a hadron, and the subprocesses are quark-antiquark annihilation and quark-gluon Compton scattering.
Actually, as always with photons, the situation is quite complex.
Together with the two different photon emission mechanisms mentioned earlier, we have four types of processes: coherent direct (coh.dir.), coherent resolved (coh.res.), incoherent direct (incoh.dir.) and incoherent resolved (incoh.res.) processes. These abbreviations will appear in many places of remained content
and we do not explain its meaning again.

The corresponding cross sections of the above four processes for dileptons photoproduction are
\begin{eqnarray}\label{photo.CS}
\sigma_{\mathrm{coh.dir.}}\!\!\!\!&=&\!\!\!\!\int dM^{2}dx_{b}dyf_{b/p}(x_{b},\mu^{2})f_{\gamma/p}(y)d\sigma(\gamma+b\rightarrow l^{+}l^{-}+b),\nonumber\\
\sigma_{\mathrm{coh.res.}}\!\!\!\!&=&\!\!\!\!\int dM^{2}dx_{b}dydx_{a'}f_{b/p}(x_{b},\mu^{2})f_{\gamma/p}(y)f_{a'/\gamma}(x_{a'},\mu^{2})\nonumber\\
\!\!\!\!&&\times d\sigma(a'+b\rightarrow l^{+}l^{-}+b),\nonumber\\
\sigma_{\mathrm{incoh.dir.}}\!\!\!\!&=&\!\!\!\!\int dM^{2}dx_{a}dx_{b}dyf_{a/p}(x_{a},\mu^{2})f_{b/p}(x_{b},\mu^{2})f_{\gamma/a}(y)\nonumber\\
\!\!\!\!&&\times d\sigma(\gamma+b\rightarrow l^{+}l^{-}+b),\nonumber\\
\sigma_{\mathrm{incoh.res.}}\!\!\!\!&=&\!\!\!\!\int dM^{2}dx_{a}dx_{b}dydx_{a'}f_{a/p}(x_{a},\mu^{2})f_{b/p}(x_{b},\mu^{2})f_{\gamma/a}(y)f_{a'/\gamma}(x_{a'},\mu^{2})\nonumber\\
\!\!\!\!&&\times d\sigma(a'+b\rightarrow l^{+}l^{-}+b),
\end{eqnarray}
where $f_{\gamma/p}(y)$ and $f_{\gamma/a}(y)$ are the photon spectrum of proton $A$ and its parton $a$, respectively.
$f_{a'/\gamma}(x_{a'},\mu^{2})$ is the parton distribution function in the resolved photon, $x_{a'}=p_{a'}/q$.
It should be emphasized that the above integrations are in the WWA form. In the calculations of exact result based on Eq.~(\ref{dabTL}),
there is no equivalent photon spectrum in Eq.~(\ref{photo.CS}), instead, $Q^{2}$ should be integrated out.

The partonic cross section of the subprocess $\gamma^{*}\rightarrow l^{+}l^{-}$ has the following form
\begin{eqnarray}\label{fgamma.incohBKH}
&&\!\!\!\!\!\!d\sigma(X+b\rightarrow l^{+}l^{-}+b)\nonumber\\
=&&\!\!\!\!\!\!\frac{\alpha_{\mathrm{em}}}{3\pi M^{2}}\sqrt{1-\frac{4m_{l}^{2}}{M^{2}}}\left(1+\frac{2m_{l}^{2}}
{M^{2}}\right)d\sigma(X+b\rightarrow \gamma^{*}+b)dM^{2},
\end{eqnarray}
where $M$ is the invariant mass of dileptons, $m_{l}$ is the lepton mass.

The fragmentation dilepton is also an important channel which involves a perturbative part - the bremsstrahlung of the virtual photon- and a nonperturbative part, described by the dilepton fragmentation function,
\begin{eqnarray}\label{f.frag}
D_{q_{c}}^{l^{+}l^{-}}(z_{c},M^{2},Q^{2})=\frac{\alpha_{\mathrm{em}}}{3\pi M^{2}}\sqrt{1-\frac{4m_{l}^{2}}{M^{2}}}\left(1+\frac{2m_{l}^{2}}{M^{2}}\right)D_{q_{c}}^{\gamma^{*}}(z_{c},Q^{2}),
\end{eqnarray}
where $D_{q_{c}}^{\gamma^{*}}(z_{c},Q^{2})$ is the virtual photon fragmentation function \cite{Kang:2008wv}, $z_{c}$ is the momentum fraction of the final state dileptons. The involved subprocesses in this channel are $q\gamma\rightarrow q\gamma$, $q\gamma\rightarrow q g$ and $\gamma g\rightarrow q\bar{q}$ for direct-photon contribution \cite{Ma:2019mwr}, and are $qq\rightarrow qq$, $qq'\rightarrow qq'$, $q\bar{q}\rightarrow q\bar{q}$, $q\bar{q}\rightarrow q'\bar{q}'$, $q\bar{q}'\rightarrow
q\bar{q}'$, $qg\rightarrow q\gamma$, $qg\rightarrow qg$ and $gg\rightarrow q\bar{q}$ for resolved-photon contribution \cite{Owens:1986mp}.

In the case of photons photoproduction, because a virtual photon can directly decay into a dilepton, the invariant cross sections of photons production can be easily derived from those of dileptons production if the invariant mass of dileptons is zero $(M^{2}=0)$.
Finally, the analytic expressions of distributions in $Q^{2}$, $y$ and $p_{T}$ can be found in \ref{Q2ypT}.

\section{Numerical results and discussion}
\label{Numerical results}
~~~~~~~~~~

We are now in a position to present our numerical results. First, several theoretical inputs need to be provided.
The mass of the proton is $m_{p}=0.938~\mathrm{GeV}$ \cite{Agashe:2014kda}, the strong coupling constant is taken as 1-loop form \cite{Ma:2015ykd}
\begin{eqnarray}\label{alfas}
&&\alpha_{\mathrm{s}}=\frac{12 \pi}{(33-2n_{f})\ln(\mu^{2}/\Lambda^{2})},
\end{eqnarray}
with $n_{f}=3$ and $\Lambda=0.2~\mathrm{GeV}$.
We adopt MMHT2014 NNLO set for the parton distribution function of proton \cite{Harland-Lang:2014zoa,Harland-Lang:2015nxa}, and choose the factorization scale to be $\mu=\sqrt{4p_{T}^{2}}$ \cite{Fu:2011zzm}.
The mass range of dileptons is $0.1\ \mathrm{GeV}<M<0.3\ \mathrm{GeV}$.
In our calculations the coherence condition is included for coherent reactions, which means that the wavelength of the photon is larger than the size of the nucleus, and the charged constituents inside the nucleus should act coherently.
This condition limits $Q^{2}$ and $y$ to very low values, $Q^{2}_{\mathrm{max}}=0.027\ \textrm{GeV}^{2}$ and $y_{\mathrm{max}}=0.16$ for proton.
Furthermore, the cross section for the LO initial parton hard scattering (hard.scat.) satisfies the following form
\begin{eqnarray}\label{hard.scat.}
\sigma_{\mathrm{hard.scat.}}=\int dM^{2}dx_{a}dx_{b}f_{a/p}(x_{b},\mu^{2})f_{b/p}(x_{b},\mu^{2})d\sigma(a+b\rightarrow l^{+}l^{-}+b),
\end{eqnarray}
where the subprocesses are $q\bar{q}\rightarrow g(\gamma^{*}\rightarrow l^{+}l^{-})$ and $qg\rightarrow q(\gamma^{*}\rightarrow l^{+}l^{-})$.

\begin{figure*}[htbp]
  \centering
  \includegraphics[width=0.45\columnwidth]{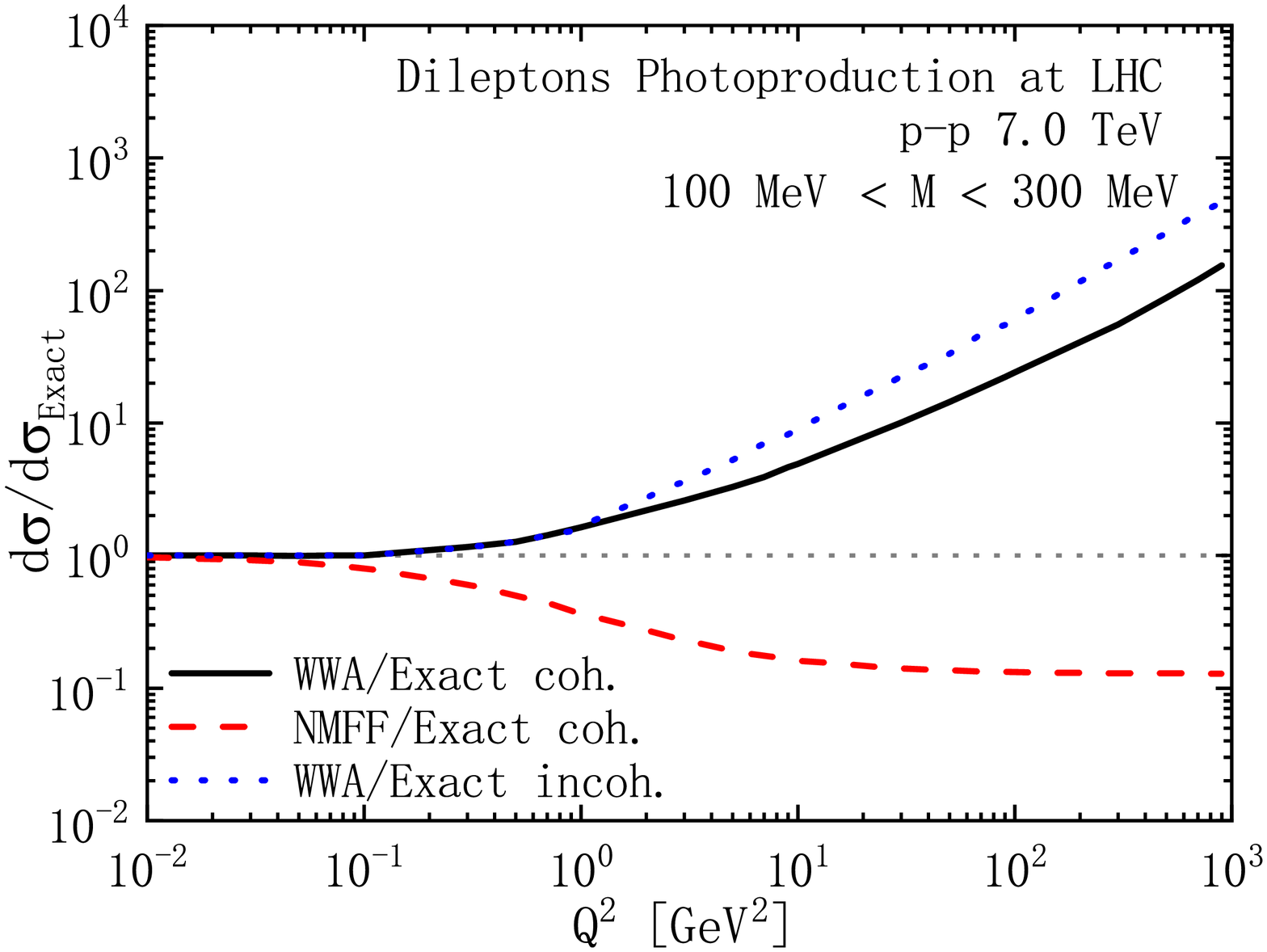}
  \includegraphics[width=0.45\columnwidth]{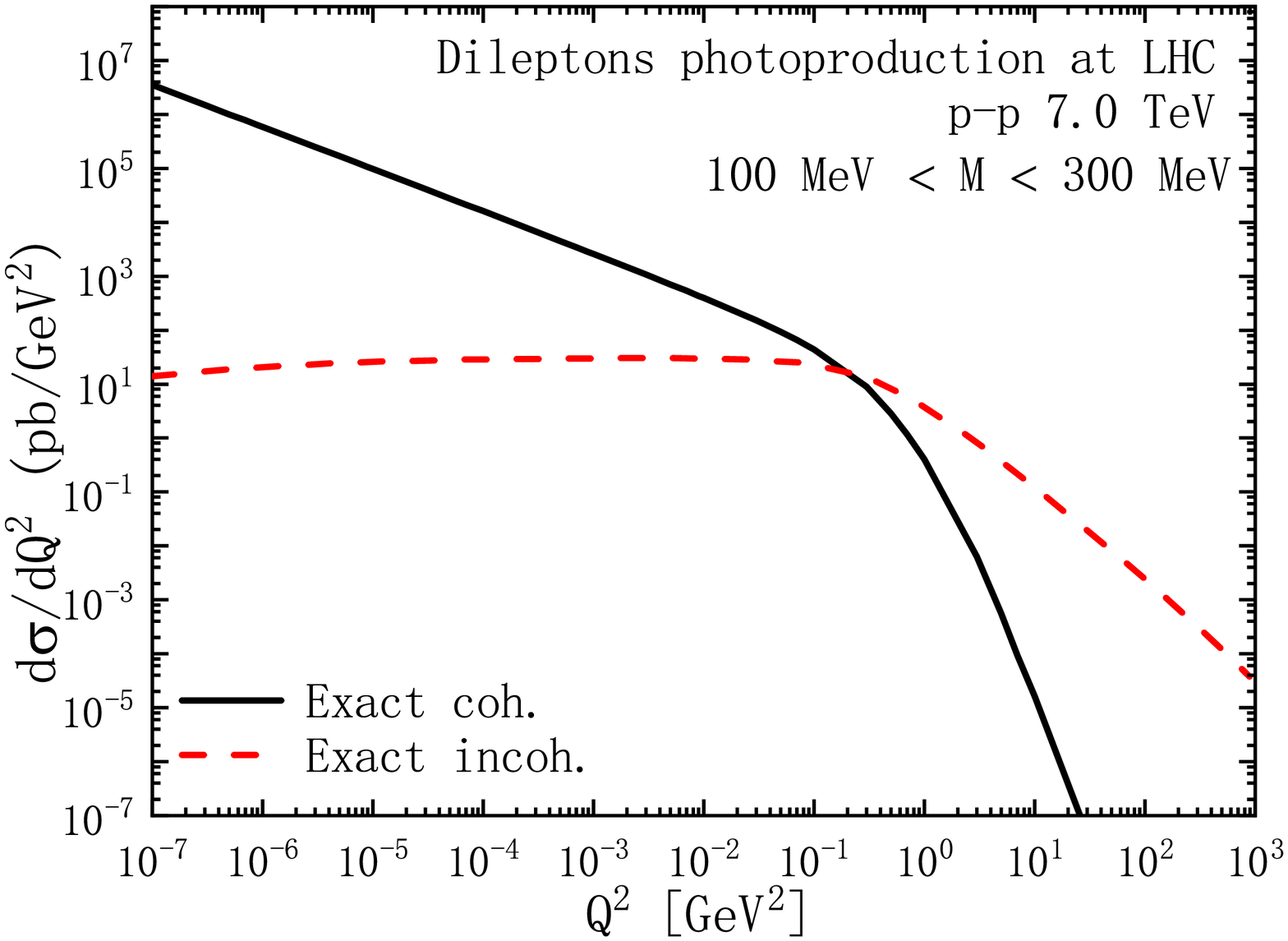}\\
  \includegraphics[width=0.45\columnwidth]{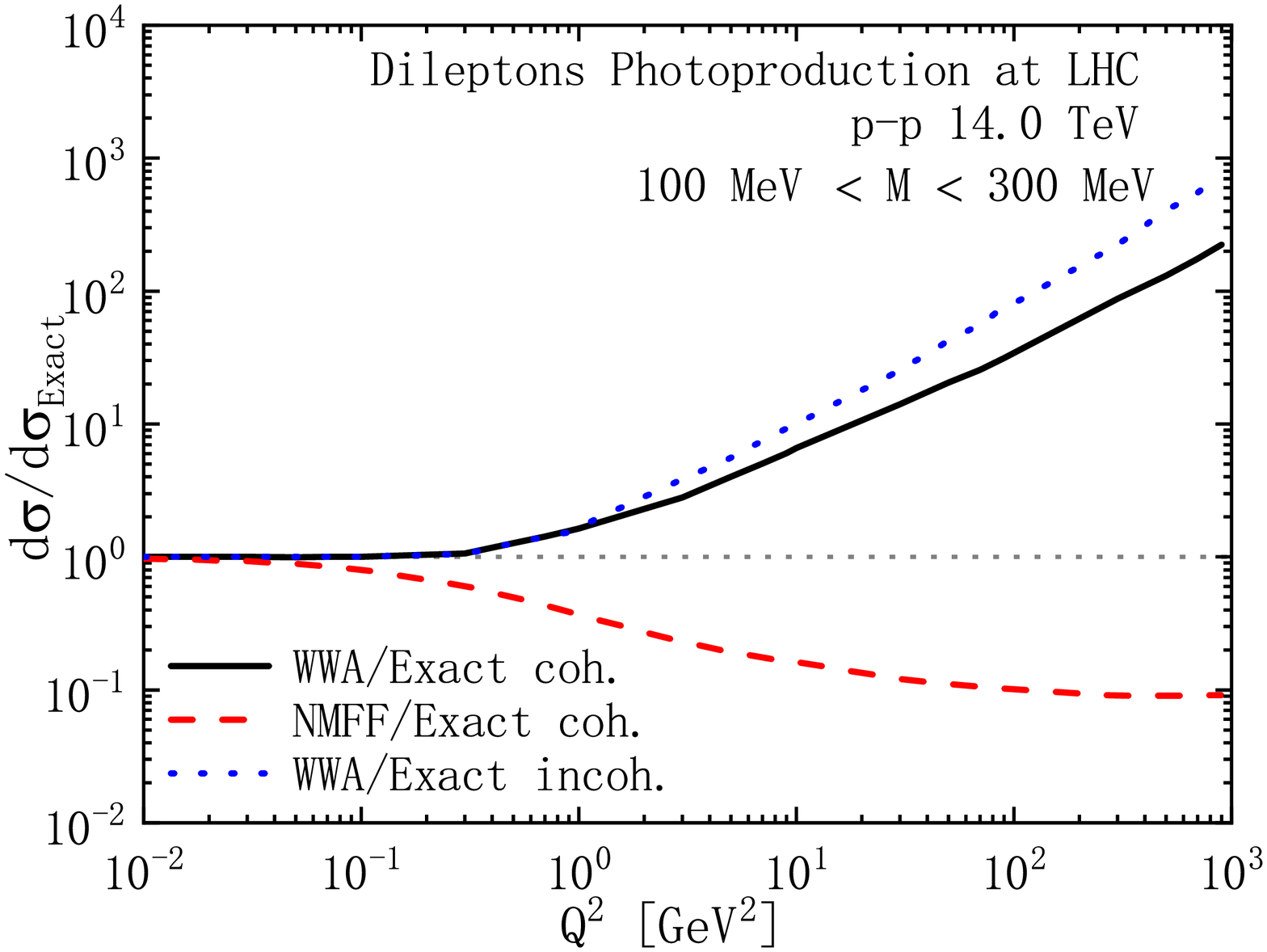}
  \includegraphics[width=0.45\columnwidth]{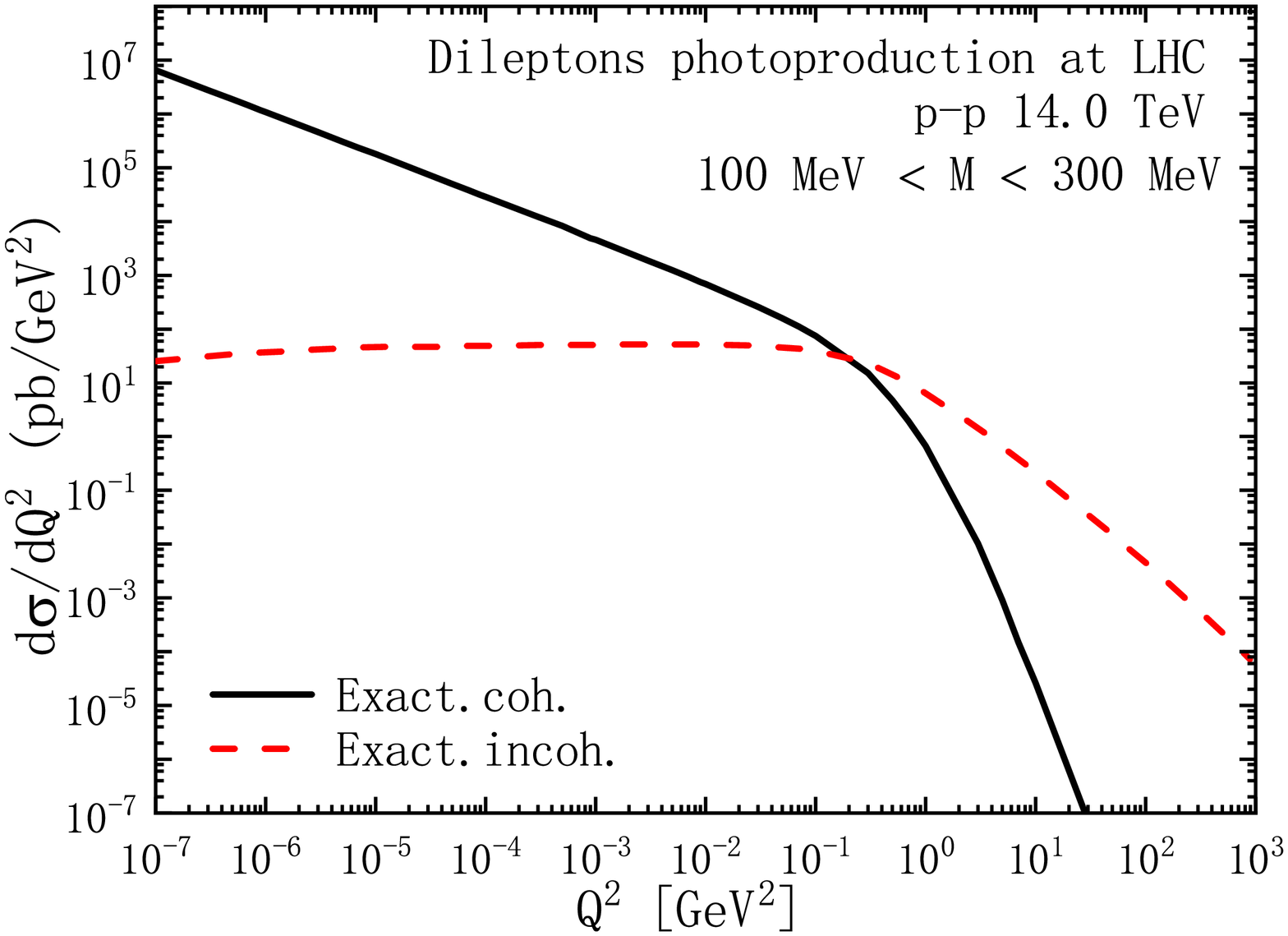}
  \caption{The left panels plot the ratios of differential cross sections in different forms to the exact ones, while the right panels plot the exact results of $d\sigma/dQ^{2}$.
  The upper and lower panels show the results in the different LHC energies, respectively.
  The abbreviation NMFF represents the exact result with no contribution of magnetic form factor.
  And each results are the sum of the direct and resolved photon contributions.}
  \label{fig:Q2.dile.}
\end{figure*}

In Fig.~\ref{fig:Q2.dile.}, we plot the comparison between differential cross sections in different forms and the exact ones in the left panels, and plot the exact results of $d\sigma/dQ^{2}$ in the right panels.
In the left panels, the WWA results nicely agree with the exact ones when $Q^{2}<0.1\ \mathrm{GeV}^{2}$, but the differences become evident with increasing $Q^{2}$.
The results that neglect the contribution of magnetic form factor (NMFF) almost have no difference compared to the exact ones when $Q^{2}<0.05\ \mathrm{GeV}^{2}$, the differences appear when $Q^{2}>0.1\ \mathrm{GeV}^{2}$ and become evident when $Q^{2}>1\ \mathrm{GeV}^{2}$.
At $Q^{2}=1\ \mathrm{GeV}^{2}$, the NMFF results deviate from the exact ones by about $63\%$; at $Q^{2}=10\ \mathrm{GeV}^{2}$ the deviations are about $84\%$.
Therefore, WWA is only valuable in small $Q^{2}$ region, its error is evident at large $Q^{2}$ domain and becomes rather serious in incoherent reactions.
And the contribution of magnetic form factor concentrates on the large $Q^{2}$ domain.
In the right panels, the coh. and incoh. contributions dominant the small and large $Q^{2}$ regions, respectively.
They become comparable at $Q^{2}=1\ \mathrm{GeV}^{2}$.
Comparing with the feature of WWA derived from the left panels, one can see that WWA is a good approximation for coherent reactions, and is essentially in contradiction with incoherent reactions.

\begin{figure*}[htbp]
  \centering
  \includegraphics[width=0.45\columnwidth]{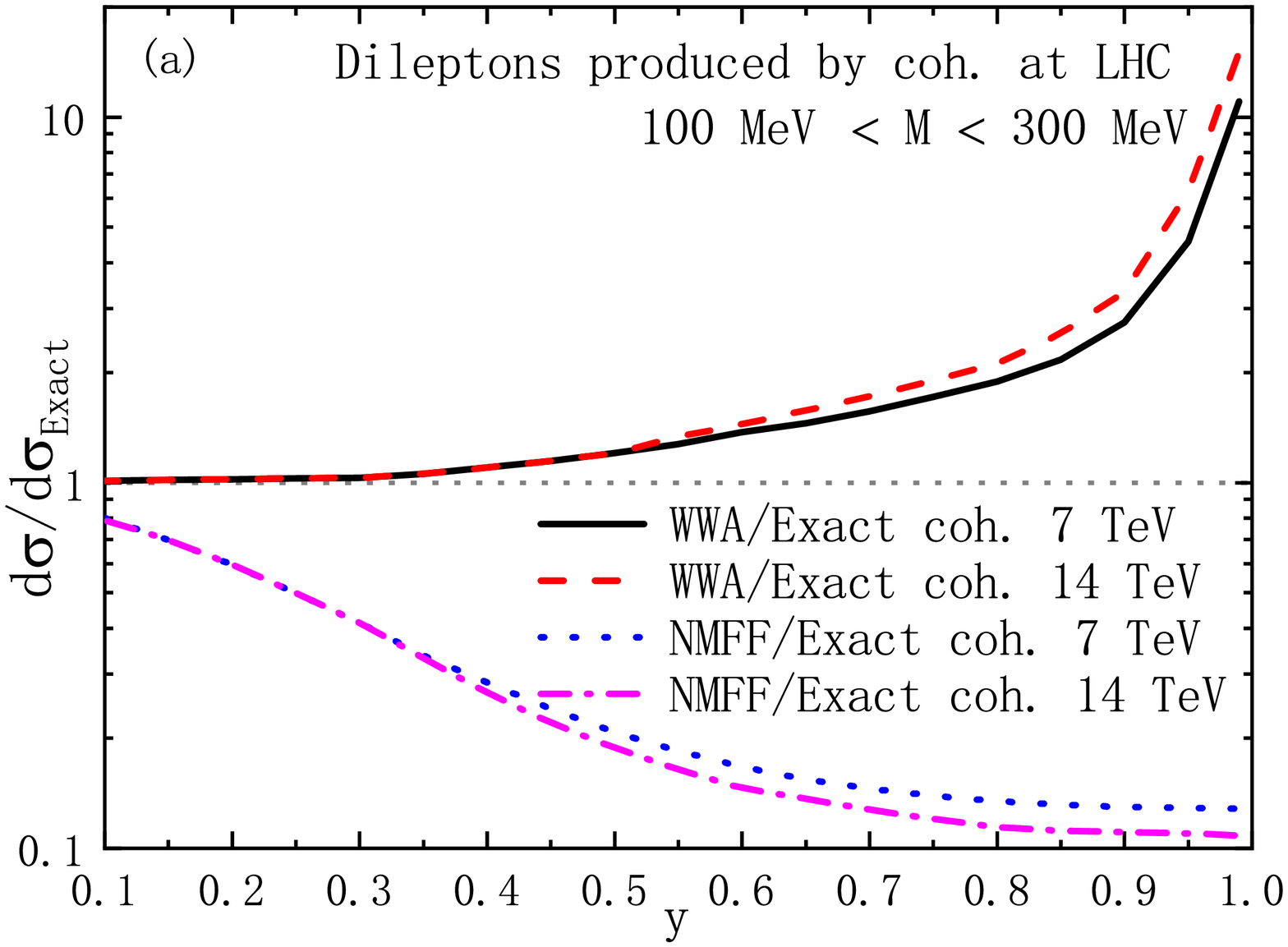}
  \includegraphics[width=0.45\columnwidth]{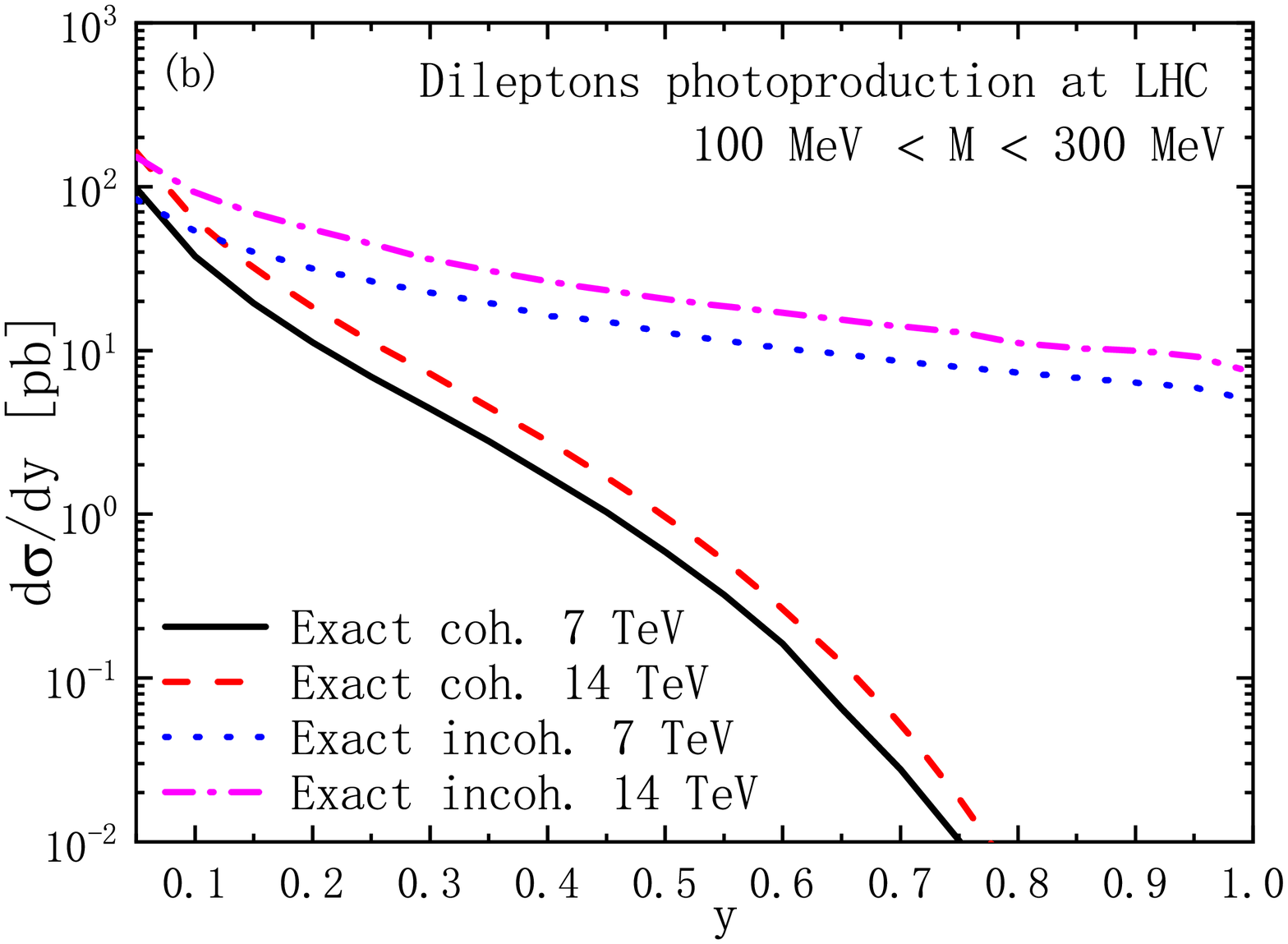}
  \caption{(a): The ratios of differential cross sections of coh. in different forms to the exact ones.
  (b): The exact results of $d\sigma/dy$.
  And each results are the sum of direct and resolved photon contributions.}
  \label{fig:y.dile.}
\end{figure*}

In Fig.~\ref{fig:y.dile.}, the results are expressed as functions of $y$.
In panel (a), the WWA results nicely agree with the exact ones when $y<0.3$, but the differences become evident with increasing $y$. Especially, when $y>0.8$ the curves show a pronounced rising.
The differences between the NMFF results and exact ones are non-negligible in the whole $y$ regions, and are the largest for large values of $y$.
At $y=0.1$, the NMFF results deviate from the exact ones by about $20\%$; at $y=0.7$ the deviations are about $86\%$.
Therefore, the error of WWA is small in small $y$ domain, while it is evident at large $y$ domain and becomes more obvious in incoh. case.
In panel (b), the curves of coh. are important when $y<0.5$ and rapidly deceased with $y$ increasing.
On the contrary, the contributions of incoh. are important in the whole $y$ regions and much higher than those of coh..

\begin{table*}[htbp]\footnotesize
\renewcommand\arraystretch{1.5}
\centering
\caption{\label{Total.CS.coh.}Total cross sections of the dilepton photoproduction in the coherent channel. }
\begin{tabular}{L{1.5cm}C{1.1cm}C{1.1cm}C{1.1cm}C{1.1cm}C{1.1cm}C{1.1cm}}
\hline
\hline
    $\sigma_{\mathrm{coh}.}$ & Exact & $f_{\mathrm{DZ}}$ & $f_{\mathrm{Ny}}$ &
    $f_{\mathrm{Kn}}$ & $f_{\mathrm{SC}}$ & $f_{\mathrm{MD}}$\\
    \hline
    $\sigma\ [\mathrm{pb}]$ $(7\ \textrm{TeV})$                    & 23.14  & 38.61  & 32.56  & 35.82  & 26.12  & 23.30 \\
    $\sigma/\sigma_{\mathrm{Exact}}$ $(7\ \textrm{TeV})$           & 1.00   & 1.67   & 1.41   & 1.55   & 1.13   & 1.01  \\
    $\sigma\ [\mathrm{pb}]$ $(14\ \textrm{TeV})$                   & 41.13  & 66.65  & 56.32  & 61.91  & 45.63  & 40.93 \\
    $\sigma/\sigma_{\mathrm{Exact}}$ $(14\ \textrm{TeV})$          & 1.00   & 1.62   & 1.37   & 1.51   & 1.11   & 1.00  \\
\hline
\hline
\end{tabular}
\end{table*}

The main purpose of the present paper is studying the features of the equivalent photon spectra which are mentioned in Section \ref{The eps}, and illustrating the advantage of Eq.~(\ref{fgamma.MD.}) that we derived.
Thus, in Tables \ref{Total.CS.coh.} and \ref{Total.CS.incoh.} we calculate the total cross sections based on the different forms of spectra.
In the case of coherent reactions, we have seen that their contributions dominate the small $Q^{2}$ and $y$ regions, which is compatible with the validity conditions of WWA.
However, the WWA errors still appear at large values of $Q^{2}$ and $y$, thus the options of upper limits, $Q^{2}_{\mathrm{max}}$ and $y_{\mathrm{max}}$, become crucial to the precision of WWA. In Table \ref{Total.CS.coh.}:
\begin{enumerate}
  \item We observe that the ratios of results based on $f_{\mathrm{DZ}}$, $f_{\mathrm{Ny}}$ and $f_{\mathrm{Kn}}$ to exact ones are about $1.4\sim1.7$, the common reason is that the integrations of these spectra are performed in the entire kinematical allowed regions: $Q^{2}_{\mathrm{max}}=\infty$ and $y_{\mathrm{max}}=1$ which include the large WWA errors, this verifies the views derived from Figs.~\ref{fig:Q2.dile.} and \ref{fig:y.dile.}, that WWA is only valuable in small $Q^{2}$ and $y$ domains.
      Actually, in most of the physically interesting cases such a dynamical cut off $\Lambda_{\gamma}$ exists such that, the WWA errors can be effectively avoided and the photo-absorption cross sections differ only slightly from their values on the mass shell.
      Thus, for the practical use of WWA, except considering the kinematically allowed regions, one should also elucidate whether there is a dynamical cut off $\Lambda^{2}_{\gamma}$, and estimate it.
      However, the definite values of $\Lambda_{\gamma}^{2}$ for different processes are essentially different, and further studies are still needed.
  \item For the spectrum $f_{\mathrm{DZ}}$, the advantage is that the form factor effects can be included, which properly describes the situation of the proton as photon emitter.
      Since the WWA is usually used in electroproduction reactions, especially in the $ep$ collision, if one obtains the spectrum of proton from that of electron by just replacing the $m_{e}$ with $m_{p}$, it would overestimate the cross section by a factor of 2 or more.
      For the spectrum $f_{\mathrm{Ny}}$, the ratios have a obvious reduction compared to those of $f_{\mathrm{DZ}}$, since $f_{\mathrm{Ny}}$ includes the $Q^{2}_{\mathrm{min}}$ term in Eq.~(\ref{fgamma.Gen.V}) which is omitted in $f_{\mathrm{DZ}}$, this factor is inversely proportional to $Q^{2}$ and thus has the obvious contribution in small $Q^{2}$ region.
      Therefore, this $Q^{2}_{\mathrm{min}}$ term can not be neglected when performing the photon spectra for coherent reactions.
      In addition, the ratios of $f_{\mathrm{Kn}}$ is somewhat higher than those of $f_{\mathrm{Ny}}$, since the effect of magnetic form factor of proton is included in this form compared to $f_{\mathrm{Ny}}$.
      We have derived in Figs. \ref{fig:Q2.dile.} and \ref{fig:y.dile.} that the contribution of magnetic form factor concentrates on the large $Q^{2}$ domain, thus this term should essentially be excluded in the coherent case.
  \item The ratios of $f_{\mathrm{SC}}$ are the smallest compared to those mentioned above, since in this form the hadronic interaction is easily excluded.
      However, we can see that the deviation from the exact results still can not be neglected.
      Finally, the results of $f_{\mathrm{MD}}$ nicely agree with the exact ones.
      Since this form has two virtues: firstly, $f_{\mathrm{MD}}$ is derived from the complete form Eq.~(\ref{fgamma.Gen.}) which properly includes the $Q^{2}_{\mathrm{min}}$ term in Eq.~(\ref{fgamma.Gen.V}) and excludes the effects of magnetic form factor;
      secondly, except considering kinematical limitations it adopts the coherence condition which limits $Q^{2}$ and $y$ to very low values, ($Q^{2}_{\mathrm{max}}=0.027\ \mathrm{GeV}^{2}$ and $y_{\mathrm{max}}=0.16$), this effectively avoid the WWA errors (which appear when $Q^{2}>0.1\ \mathrm{GeV}^{2}$ and $y>0.3$ [Figs. \ref{fig:Q2.dile.}, \ref{fig:y.dile.}]).
\end{enumerate}

\begin{table*}[htbp]\footnotesize
\renewcommand\arraystretch{1.5}
\centering
\caption{\label{Total.CS.incoh.}Total cross sections of the dilepton photoproduction in the incoherent channel.
The results of WWA are deduced from Eq.~(\ref{dWWA.Gen.}).}
\begin{tabular}{L{1.5cm}C{1.5cm}C{1.5cm}C{1.5cm}C{1.5cm}}
\hline
\hline
    $\sigma_{\mathrm{incoh}.}$   &  Exact  & WWA  & $f_{\mathrm{incoh.}}$ [Eq. (\ref{fgamma.incohI.})]  &  $f_{\mathrm{BKT}}$ \\
\hline
  $\sigma\ [\mathrm{pb}]$ $(7\ \textrm{TeV})$           & 21.87  & 93.25   & 94.10   & 147.80 \\
  $\sigma/\sigma_{\mathrm{Exact}}$ $(7\ \textrm{TeV})$  & 1.00   & 4.26    & 4.30    & 6.76   \\
  $\sigma\ [\mathrm{pb}]$ $(14\ \textrm{TeV})$          & 35.14  & 170.11  & 175.27  & 266.04 \\
  $\sigma/\sigma_{\mathrm{Exact}}$ $(14\ \textrm{TeV})$ & 1.00   & 4.84    & 4.99    & 7.57   \\
\hline
\hline
\end{tabular}
\end{table*}

For the incoherent reactions, we have seen that its contribution dominates the large $Q^{2}$ region and almost the whole $y$ region, this is in contradiction with the validity conditions of WWA.
In Table \ref{Total.CS.incoh.}, we observe that the ratios of the WWA parameterizations to the exact ones are prominent compared to those of the coherent cases.
This quantitatively verifies the inapplicability of WWA in incoherent reactions.
For the spectrum $f_{\mathrm{incoh}}$, the ratios should be actually much higher than the values given in Table \ref{Total.CS.incoh.}, since the term of weighting factor $1-G_{\mathrm{E}}^{2}(Q^{2})$ is omitted in this form, and will lead to unreasonable divergency in small $Q^{2}$ region. 
This unphysical results are essentially caused by the serious double counting (The detailed discussion can be found in Ref.~\cite{Ma:2018zzq}). 
However, an artificial cutoff $Q^{2}_{\mathrm{min}}=1\ \mathrm{GeV}^{2}$ is used in Eq.~(\ref{fgamma.incohI.}) to avoid this divergency.
But we can see that the result is still not accurate.
Furthermore, the ratios of $f_{\mathrm{BKT}}$ are the largest, since $f_{\mathrm{BKT}}$ is originally derived from $ep$ scattering, but is directly expanded to describe the probability of finding a photon in any relativistic fermion and to deal with hadronic collisions in Refs.~\cite{Kniehl:1990iv,Yu:2015kva,Yu:2017pot}, this will overestimate the cross sections.
Therefore, the accurate expression Eq.~(\ref{dabTL}) should be applied for the incoherent photon emission.

\begin{figure*}[htbp]
  \centering
  \includegraphics[width=0.45\columnwidth]{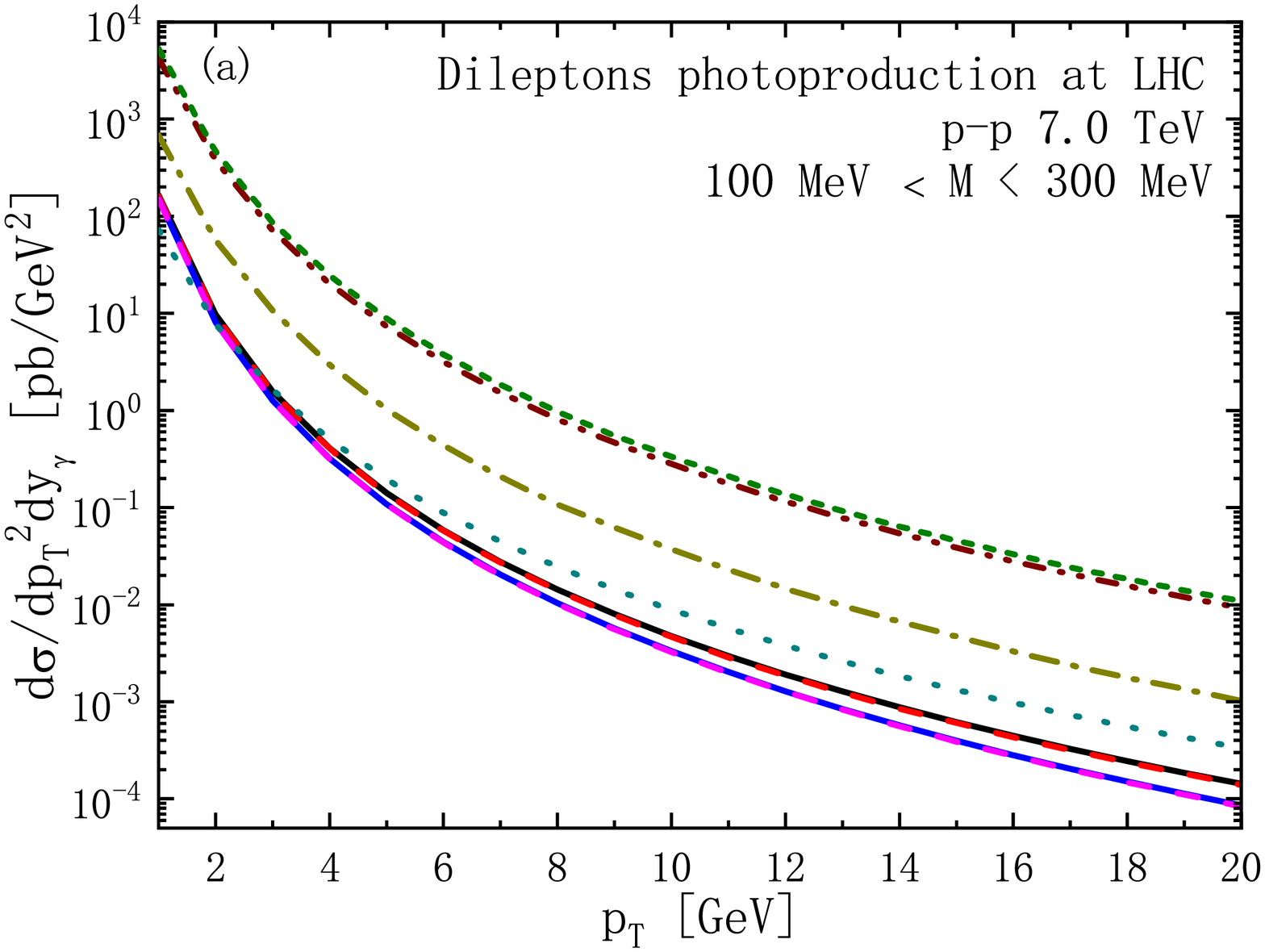}
  \includegraphics[width=0.45\columnwidth]{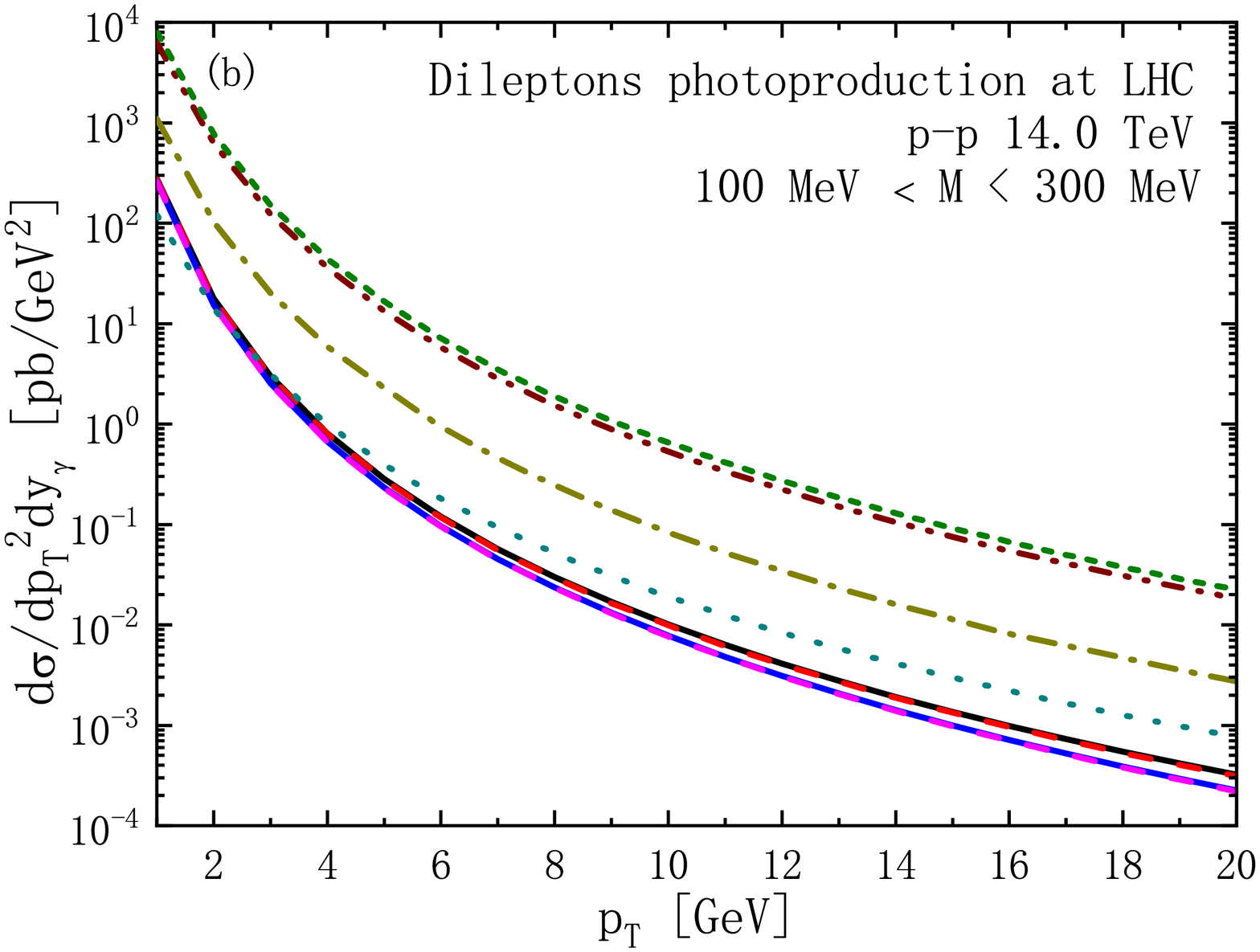}
  \caption{Black solid line---the exact result of coherent reactions [coh.(dir.+res.)].
  Red dash line---the WWA result based on $f_{\mathrm{MD}}$ [coh.(dir.+res.)].
  Blue solid line---the exact result of fragmentation dileptons produced by coherent reactions [coh.(dir.+res.)-frag.].
  Magenta dash line---the WWA result based on $f_{\mathrm{MD}}$ for fragmentation dilepton photoproduction [coh.(dir.+res.)-frag.].
  Dark cyan dot line---the exact result of incoherent reactions [incoh.(dir.+res.)].
  Dark yellow dash dot line---the exact result of fragmentation dileptons produced by incoherent reactions [incoh.(dir.+res.)-frag.].
  Wine dash dot dot line---the initial parton hard scattering processes [hard.scat.].
  Olive short dash line---the sum of the above processes.
  The solid lines coincide with the dash lines in the whole $p_{T}$ domain.
   }
  \label{fig:pT.dile.}
\end{figure*}

In Fig.~\ref{fig:pT.dile.}, we adopt the complete form Eq.~(\ref{dabTL}) and the modified equivalent photon spectrum Eq.~(\ref{fgamma.MD.}) to plot the dileptons photoproduction in $p_{T}$ distribution.
We find that the exact results of coherent photon emission are consistent with the ones of modified equivalent photon spectrum Eq.~(\ref{fgamma.MD.}) in the whole $p_{T}$ regions, this verifies again the virtue of $f_{\mathrm{MD}}$.
Since WWA is inapplicable for incoherent reactions, we only present the exact results.
One can see that the contributions of coh. and incoh. are comparable with each other, and are about two orders of magnitudes (OOM) smaller than the LO initial parton hard scattering.
This is similar to the results in Ref.\cite{Fu:2011zzm}, but is very different from the results in Ref.~\cite{Yu:2017pot} where the incoherent contributions are about two OOMs larger than the coherent ones, and the contributions of photoproduction processes are also about two OOMs larger than LO hard.scat.
Finally, it can be found that the correction of photoproduction processes to the dileptons production is about $20\%$.

\begin{figure*}[htbp]
  \centering
  \includegraphics[width=0.45\columnwidth]{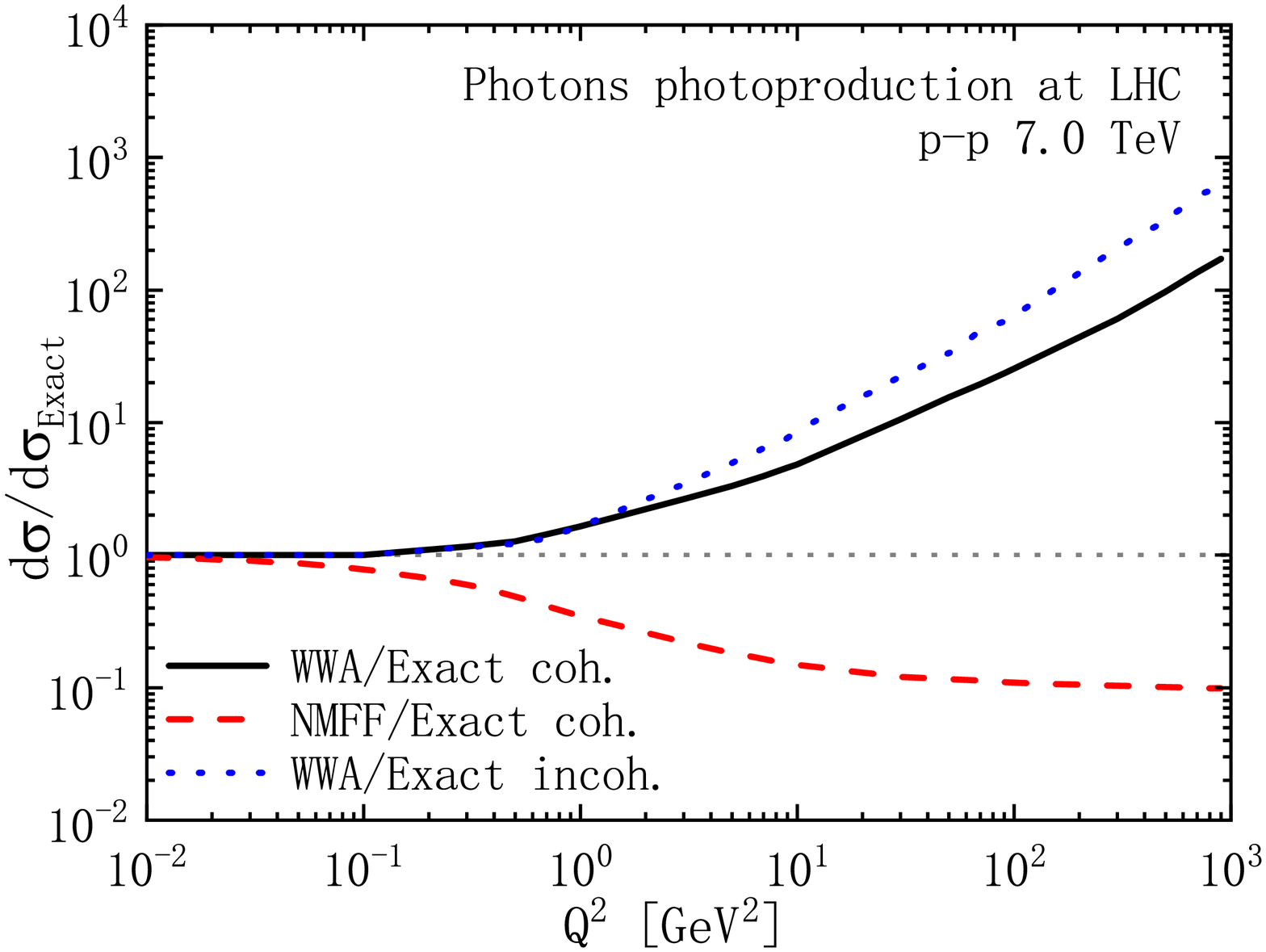}
  \includegraphics[width=0.45\columnwidth]{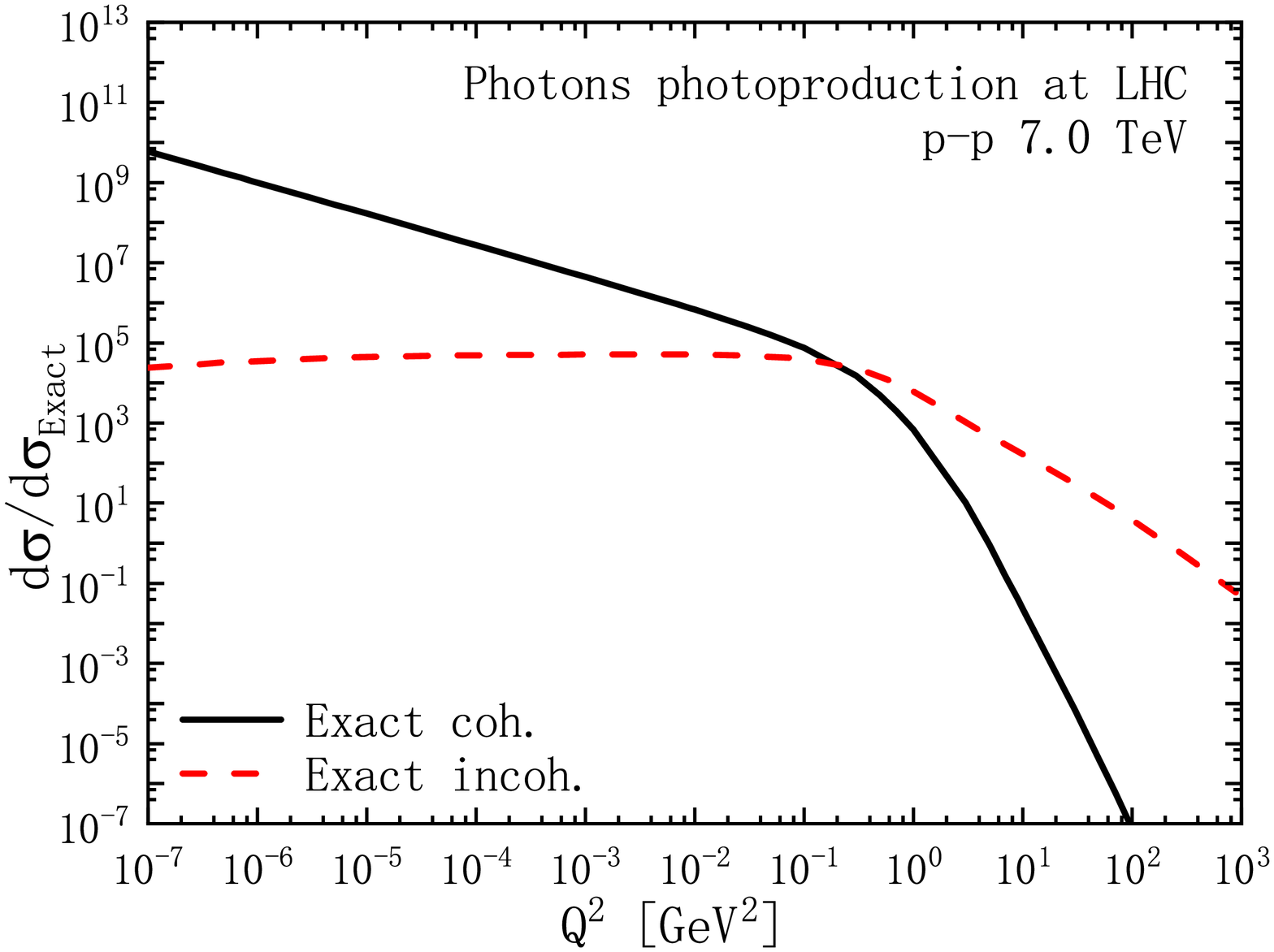}\\
  \includegraphics[width=0.45\columnwidth]{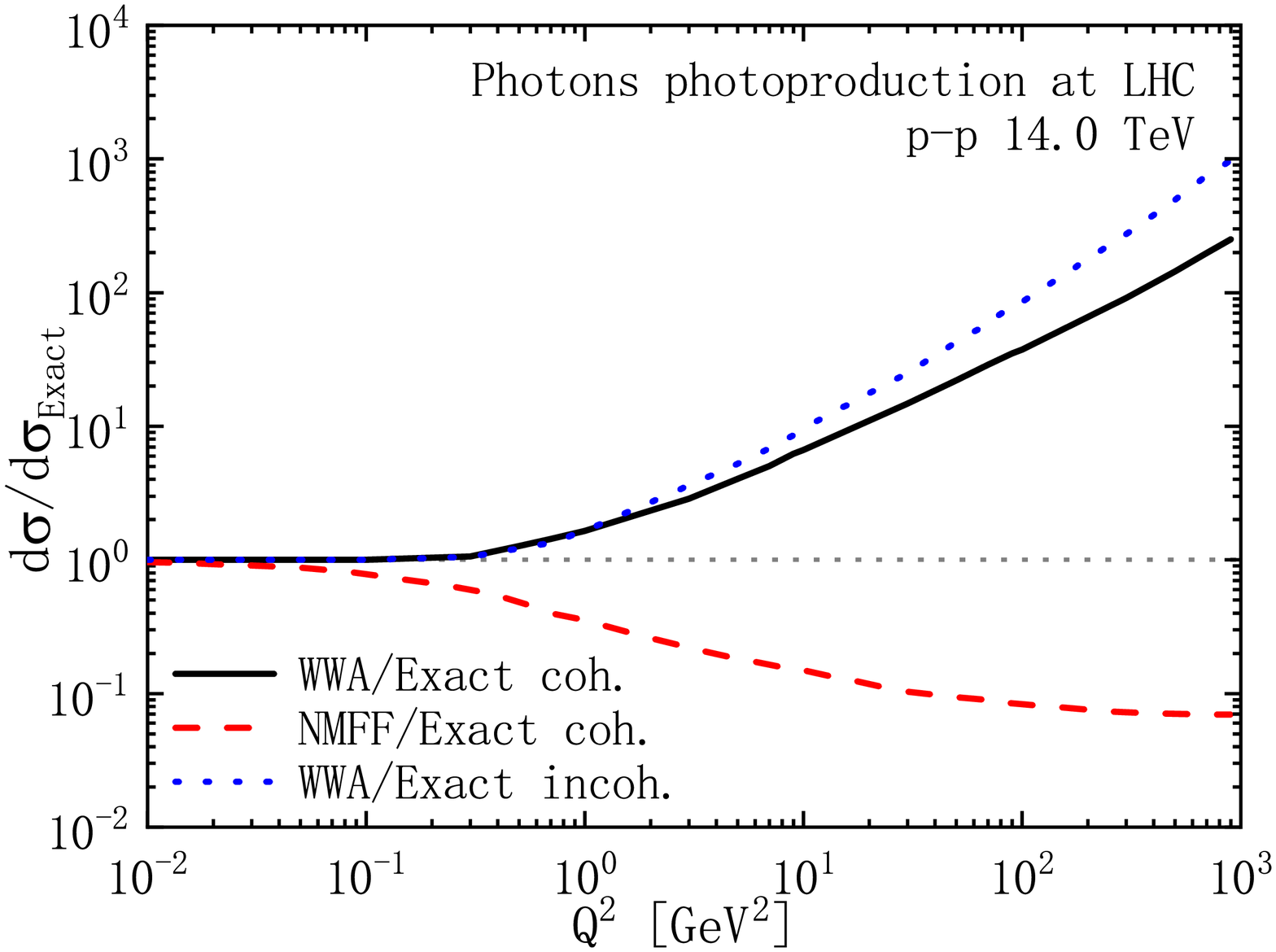}
  \includegraphics[width=0.45\columnwidth]{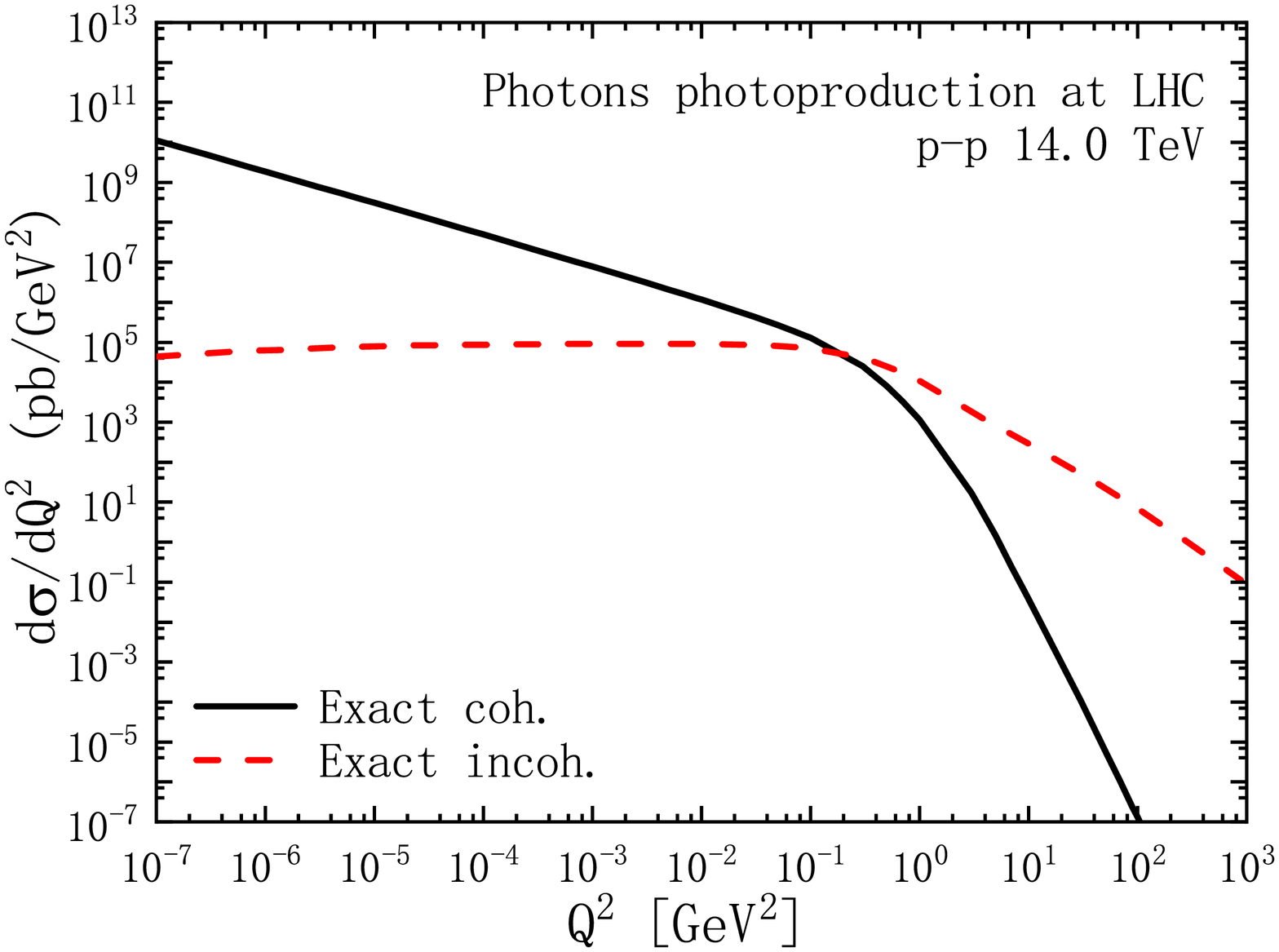}
  \caption{Same as Fig.~\ref{fig:Q2.dile.} but for photons.}
  \label{fig:Q2.pho.}
\end{figure*}

\begin{figure*}[htbp]
  \centering
  \includegraphics[width=0.45\columnwidth]{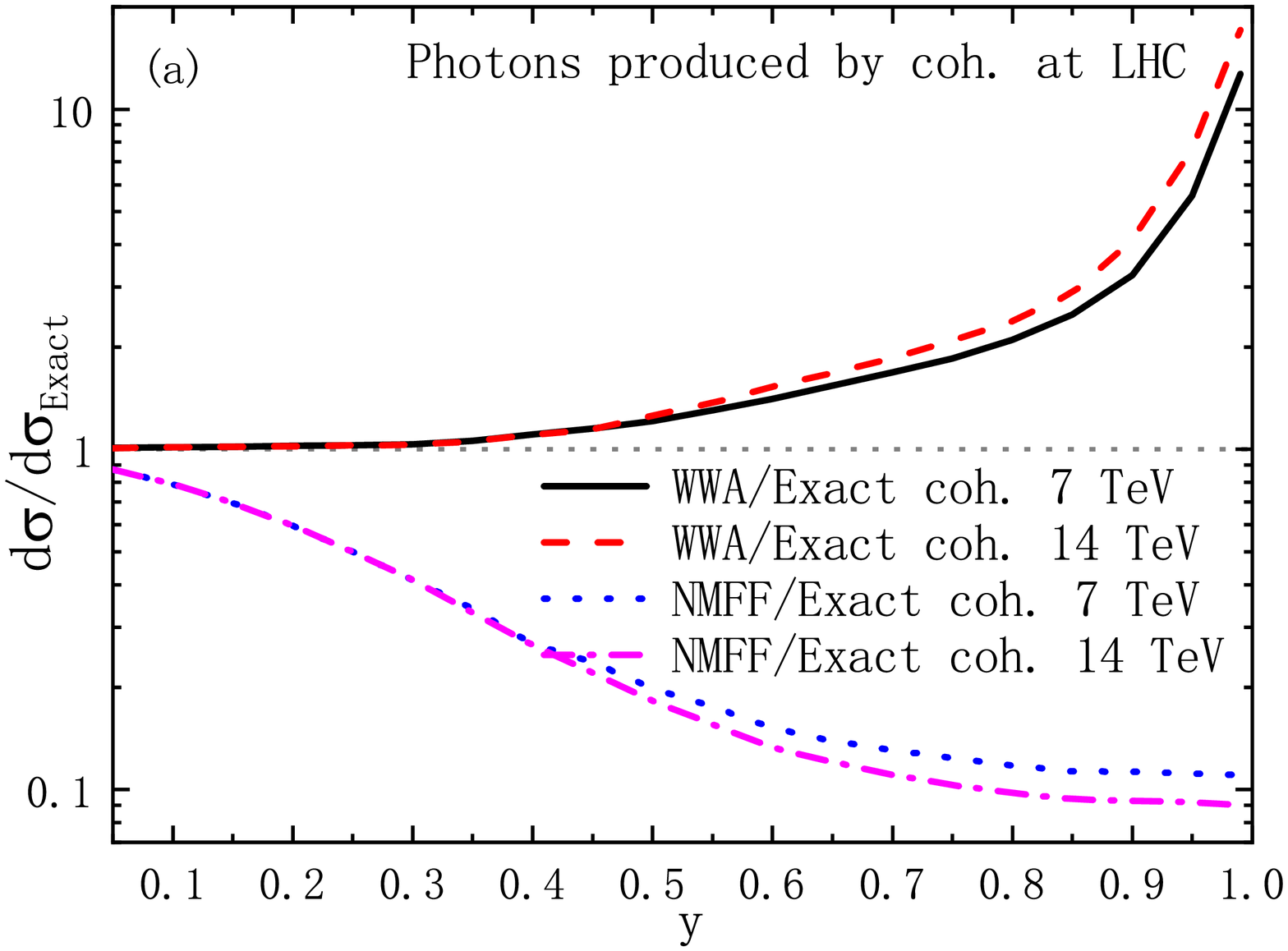}
  \includegraphics[width=0.45\columnwidth]{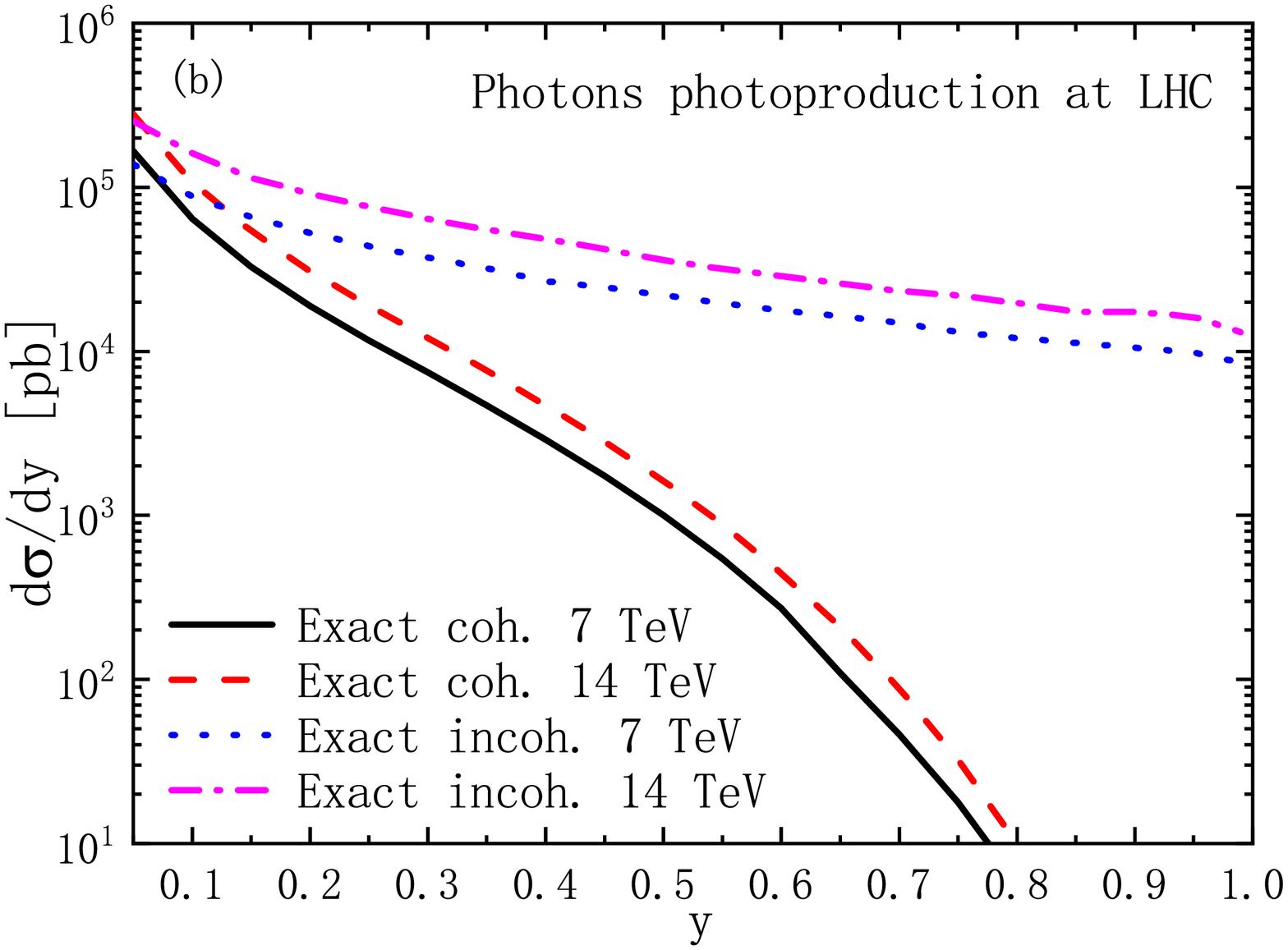}
  \caption{Same as Fig.~\ref{fig:y.dile.} but for photons.}
  \label{fig:y.pho.}
\end{figure*}

\begin{table*}[htbp]\footnotesize
\renewcommand\arraystretch{1.5}
\centering
\caption{\label{Total.CS.coh.pho.}Total cross sections of the photoproduction of photon in the coherent channel. }
\begin{tabular}{L{1.5cm}C{1.1cm}C{1.1cm}C{1.1cm}C{1.1cm}C{1.1cm}C{1.1cm}}
\hline
\hline
    $\sigma_{\mathrm{coh}.}$ & Exact & $f_{\mathrm{DZ}}$ & $f_{\mathrm{Ny}}$ &
    $f_{\mathrm{Kn}}$ & $f_{\mathrm{SC}}$ & $f_{\mathrm{MD}}$\\
    \hline
    $\sigma\ [\mathrm{nb}]$ $(7\ \textrm{TeV})$                    & 39.56  & 66.04  & 55.65  & 61.27  & 44.69  & 39.83 \\
    $\sigma/\sigma_{\mathrm{Exact}}$ $(7\ \textrm{TeV})$           & 1.00   & 1.67   & 1.41   & 1.55   & 1.13   & 1.01  \\
    $\sigma\ [\mathrm{nb}]$ $(14\ \textrm{TeV})$                   & 70.35  & 113.65 & 96.41  & 105.57 & 79.04  & 70.65 \\
    $\sigma/\sigma_{\mathrm{Exact}}$ $(14\ \textrm{TeV})$          & 1.00   & 1.62   & 1.37   & 1.50   & 1.12   & 1.00  \\
\hline
\hline
\end{tabular}
\end{table*}

\begin{table*}[htbp]\footnotesize
\renewcommand\arraystretch{1.5}
\centering
\caption{\label{Total.CS.incoh.pho.}Total cross sections of the photoproduction of photon in the incoherent channel.
The results of WWA are deduced from Eq. (\ref{dWWA.Gen.}).}
\begin{tabular}{L{1.5cm}C{1.5cm}C{1.5cm}C{1.5cm}C{1.5cm}}
\hline
\hline
    $\sigma_{\mathrm{incoh}.}$   &  Exact  & WWA  & $f_{\mathrm{incoh.}}$ [Eq. (\ref{fgamma.incohI.})]  &  $f_{\mathrm{BKT}}$ \\
\hline
  $\sigma\ [\mathrm{nb}]$ $(7\ \textrm{TeV})$           & 37.89  & 158.67  & 160.41  & 252.06 \\
  $\sigma/\sigma_{\mathrm{Exact}}$ $(7\ \textrm{TeV})$  & 1.00   & 4.19    & 4.23    & 6.65   \\
  $\sigma\ [\mathrm{nb}]$ $(14\ \textrm{TeV})$          & 61.54  & 291.71  & 311.17  & 474.15 \\
  $\sigma/\sigma_{\mathrm{Exact}}$ $(14\ \textrm{TeV})$ & 1.00   & 4.74    & 5.06    & 7.70   \\
\hline
\hline
\end{tabular}
\end{table*}

\begin{figure*}[htbp]
  \centering
  \includegraphics[width=0.45\columnwidth]{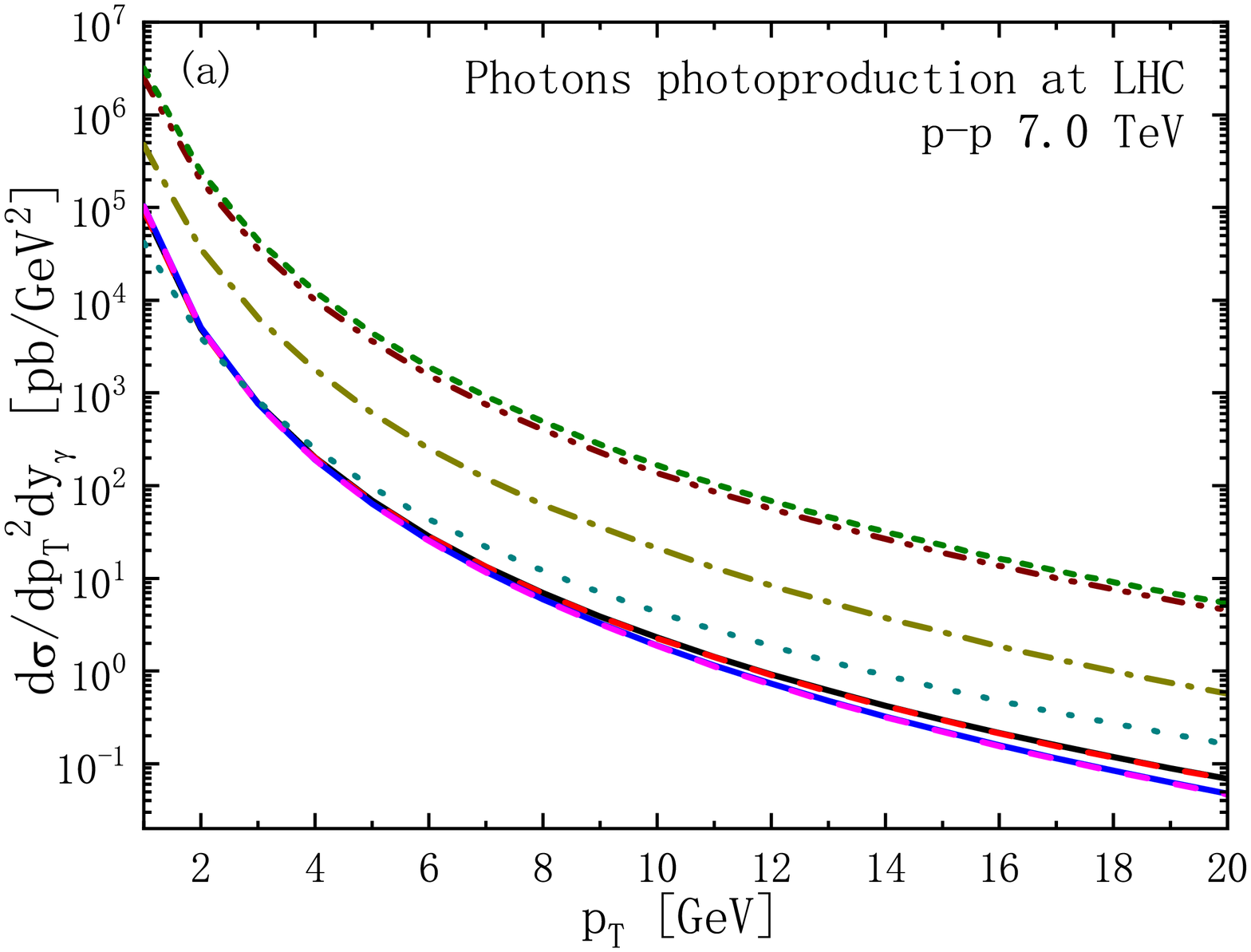}
  \includegraphics[width=0.45\columnwidth]{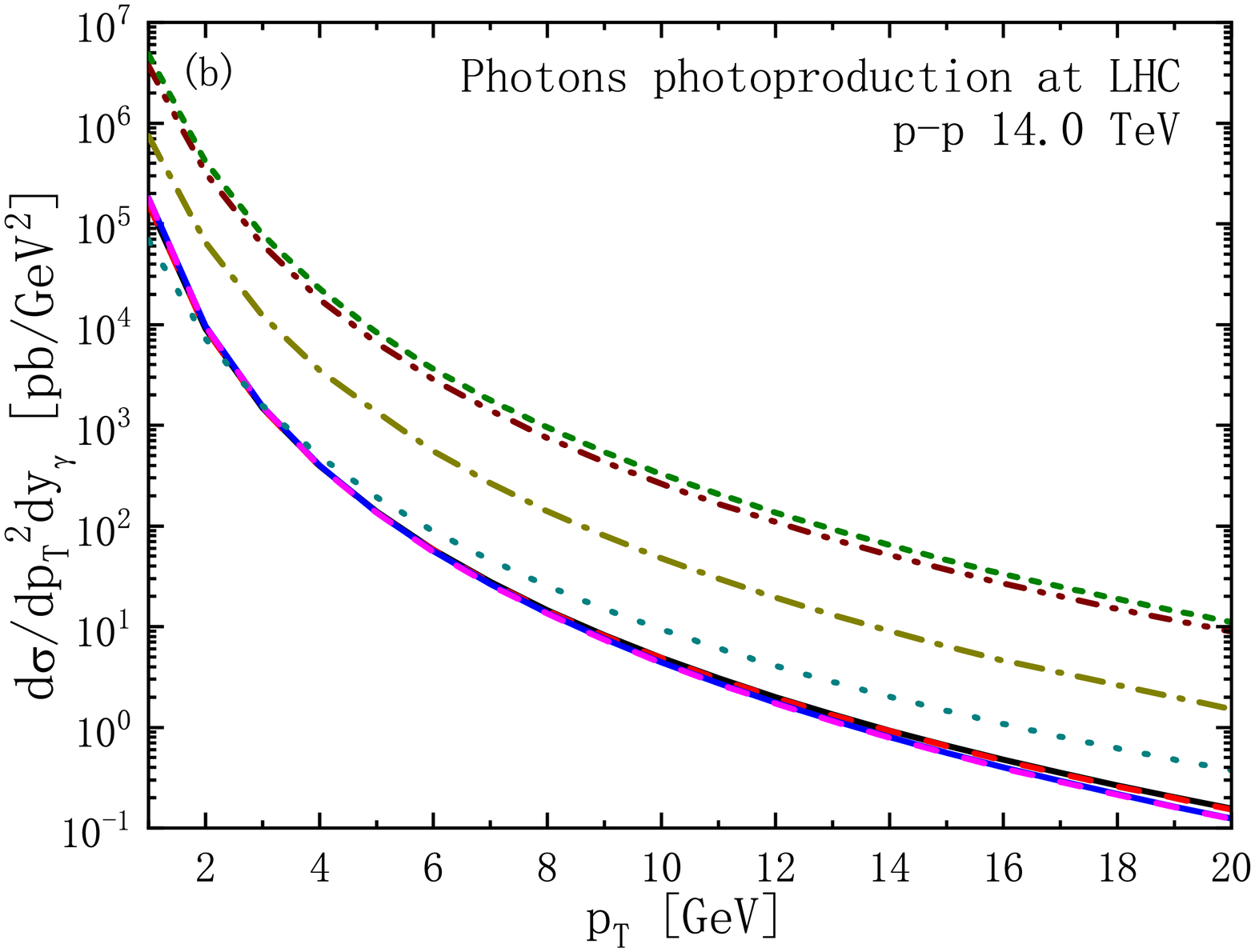}
  \caption{Same as Fig.~\ref{fig:pT.dile.} but for photons.}
  \label{fig:pT.pho.}
\end{figure*}

Figs.~\ref{fig:Q2.pho.} and \ref{fig:y.pho.} present the $Q^{2}$ and $y$ distributions of real photons photoproduction.
It is shown that the differences between the WWA results and exact ones are more obvious.
The total cross sections of real photons are given by Tables \ref{Total.CS.coh.pho.} and \ref{Total.CS.incoh.pho.}, the differences between the exact results and the
ones of equivalent photon spectra are still prominent.
And Eq.~(\ref{fgamma.MD.}) is also a good approximation for the coherent photoproduction of real photons.
In Fig.~\ref{fig:pT.pho.}, the $p_{T}$ distribution of real photons production is plotted.
We also compare our results of real photons to Refs.\cite{Fu:2011zzm,Yu:2017pot}, where the inaccuracies of equivalent photon spectra are more evident.
~~~~~~~~~~

\section{Conclusion}
\label{Conclusion}
~~~~~~~~~~

In this work, we have studied the validity of several equivalent photon spectra which are widely used in previous studies and derived a modified photon flux of proton, by giving a consistent analysis of the terms neglected in going from the accurate expression to the WWA one.
Since the equivalent photon spectrum plays the fundamental role in the photoproduction processes, we have taken photoproductions of photons and dileptons as examples, to express the comparison between the WWA results and the exact ones as the distributions in $Q^{2}$ and $y$, the total cross sections were also given.
The modified equivalent photon spectrum reproduces the exact result within less than one percent.
In the sequel, we have applied the accurate expression and the modified photon flux to the $p_{T}$ dependent cross sections.
Our results are different from the results in a previous study, and show that the corrections of photoproduction processes to the dileptons and photons prolductions are about $20\%$.
~~~~~~~~~~

\section*{Acknowledgements}
~~~~~~~

This work is supported in part by National Key R \& D Program of China under grant No. 2018YFA0404204, the NSFC (China) grant 11575043.
Z. M. is supported by the fellowship of China Postdoctoral Science Foundation under grant No. 2021M692729, and by Yunnan Provincial New Academic Researcher Award for Doctoral Candidates.
~~~~~~~~~~

\appendix
\section{The $Q^{2}$, $y$ and $p_{T}$ distributions}
\label{Q2ypT}
~~~~~~~~~~

It is straightforward to obtain the distributions in $Q^{2}$, $y$ and $p_{T}$ by accordingly reordering and redefining the involved integration variables.
To obtain the $Q^{2}$ and $y$ distributions, it is convenient to do the calculations in the rest frame of $\alpha$, where $|\textbf{q}|=|\textbf{p}_{\alpha'}|=r$, $Q^{2}=-q^{2}=(p_{\alpha}-p_{\alpha'})^{2}=2m_{\alpha}(\sqrt{r^{2}+m_{\alpha}^{2}}-m_{\alpha})$,
$d^{3}p_{\alpha}'=r^{2}drd\cos\theta d\varphi$, and $y=(q_{0}-|\textbf{p}_{\beta}|r\cos\theta/E_{\beta})/m_{\alpha}$.
By using the Jacobian determinant,
\begin{eqnarray}\label{Jac.Q2.y}
dydQ^{2}=\left|\frac{D(Q^{2},y)}{D(\cos\theta,r)}\right|d\cos\theta dr=\frac{2|\textbf{p}_{b}|r^{2}}{E_{\alpha'}E_{b}}d\cos\theta dr,
\end{eqnarray}
where $E_{b}=(s_{\alpha b}-m_{\alpha}^{2}-m_{b}^{2})/2m_{\alpha}$, the phase-space element $d^{3}p'_{\alpha}/(2\pi)^{3}2E'_{\alpha}$ in Eq.~(\ref{dab.Gen2}) can be changed accordingly
\begin{eqnarray}\label{phaseQ2y}
\frac{d^{3}p'_{\alpha}}{(2\pi)^{3}2E'_{\alpha}}=\frac{1}{4(2\pi)^{3}}\frac{E_{b}}{|\textbf{p}_{b}|}dydQ^{2}d\varphi
\end{eqnarray}

For obtaining the $p_{T}$ distribution, the variables $x_{b}$ and $\hat{t}$ should be transformed into the following form in the case coh.dir. by using the Jacobian determinant
\begin{eqnarray}\label{Jac}
d\hat{t}dx_{b}=\mathcal{J}_{\mathrm{coh.dir.}}dy_{r}dp_{T}=\left|\frac{D(x_{b},\hat{t})}{D(p_{T},y_{r})}\right|dy_{r}dp_{T},
\end{eqnarray}
and the Jacobian determinant for the rest processes are listed as follows
\begin{eqnarray}\label{Jac.rel.}
\mathrm{incoh.dir.}\!\!\!\!&:&\!\!\!\!~~d\hat{t}dx_{b}=\frac{\mathcal{J}_{\mathrm{coh.dir.}}}{x_{a}}dy_{r}dp_{T},\nonumber\\
\mathrm{coh.res.}\!\!\!\!&:&\!\!\!\!~~d\hat{t}dz_{a'}=\frac{\mathcal{J}_{\mathrm{coh.dir.}}}{x_{b}}dy_{r}dp_{T},\nonumber\\
\mathrm{incoh.res.}\!\!\!\!&:&\!\!\!\!~~d\hat{t}dz_{a'}=\frac{\mathcal{J}_{\mathrm{coh.dir.}}}{x_{a}x_{b}}dy_{r}dp_{T}.
\end{eqnarray}
the detailed expressions can be found in Ref.\cite{Ma:2021jes}.
For fragmentation processes,  the variables $z_{c}$ and $\hat{t}$ should be transformed into the following form
\begin{eqnarray}\label{Jac.frag.}
d\hat{t}dz_{c}=\mathcal{J}dy_{r}dp_{T}=\left|\frac{D(z_{c},\hat{t})}{D(p_{T},y_{r})}\right|dy_{r}dp_{T}.
\end{eqnarray}




\end{document}